\documentclass[a4paper,11pt]{article}
\usepackage{jmehppub}

\usepackage[T1]{fontenc} 

\usepackage{setspace}
\usepackage{subcaption}
\usepackage{physics}
\usepackage{dsfont}
\usepackage{tensor}
\usepackage[normalem]{ulem}
\usepackage{xcolor}
\usepackage{graphicx}
\usepackage[export]{adjustbox}
\usepackage{mathtools}
\usepackage{mathbbol}
\usepackage[centertableaux]{ytableau}
\usepackage{sidenotes}
\usepackage{mathdots}
\usepackage{tikz}
\usepackage{comment}

\colorlet{darkgreen}{green!50!black}

\newcommand{\be}{\begin{equation}}
\newcommand{\ee}{\end{equation}}
\newcommand{\beq}{\begin{equation}}
\newcommand{\eeq}{\end{equation}}

\newcommand{\mc}[1]{\mathcal{#1}}
\newcommand{\ms}[1]{\mathsf{#1}}

\newcommand{\bmn}{\text{BMN}}
\newcommand{\pa}{\partial}

\usepackage{tikz,tikz-3dplot,xcolor}
\usepackage[table]{xcolor}
\usepackage[dvipsnames]{xcolor}
\usetikzlibrary{knots,hobby,calc,decorations.markings,decorations.pathmorphing,decorations.pathreplacing,fadings,patterns,perspective,arrows.meta,shapes.misc}
\tikzfading[name=middle,
            top color=transparent!100,
            bottom color=transparent!100,
            right color=transparent!100,
            left color=transparent!100,
            middle color=transparent!0]
\tikzset{
	partial ellipse/.style args={#1:#2:#3}{
		insert path={+ (#1:#3) arc (#1:#2:#3)}
	}
}
\tikzset{
  every overlay node/.style={
    draw=black,fill=white,rounded corners,anchor=north west,
  },
}
\tikzfading
[
  name=fade out,
  inner color=transparent!0,
  outer color=transparent!100
]

\title{Signatures of Bulk Topology from the `t Hooft Worldsheet}
    \author[a]{Jackson R. Fliss,}
    \author[b]{ Alexander Frenkel}

    \affiliation[a]{Physique Th\'eoretique et Math\'ematique, Universit\'e Libre de Bruxelles \& International Solvay Institutes, CP 231, 1050 Bruxelles, BE}
    \affiliation[b] {Simons Center for Geometry and Physics, Stony Brook University, Stony Brook, NY 11794, USA}

   \emailAdd{jackson.fliss@ulb.be}
   \emailAdd{afrenkel@scgp.stonybrook.edu}

   \abstract{
         We study entanglement entropy in a certain ansatz of matrix quantum mechanics (MQM) wavefunctions which behave as generating functionals for a planar `t Hooft expansion. We argue that these `t Hooft diagrammatics can directly probe aspects of the topology of the bulk dual, such as the contractibility of a thermal or replica $S^1$ cycle. Our argument relies on treating `t Hooft diagrams as a non-perturbative definition of off-shell string theory worldsheets in which contractibility of replica cycle leads to worldsheets punctured by a conical deficit (as first described by Susskind and Uglum). We describe how a deconfinement transition in the MQM is tied to the exchange of dominant topologies, as well as to a candidate description of open string `edge modes' of Susskind and Uglum. Lastly, we study toy models of black hole evaporation and show that the boundary planar expansion itself diagnoses the exchange of dominance between the disconnected topology
         and replica wormhole topology, without ensemble averaging and without independently postulating which bulk saddles should be included.
   }
\date{\today}

\begin{document}

\maketitle

\section{Introduction}\label{sec:intro}

Contributions from competing topologies in the gravitational path integral have led to a host of exciting progress in our understanding of quantum gravity \cite{Penington:2019kki,Almheiri:2019qdq,Saad:2021uzi,Usatyuk:2024isz,Harlow:2025pvj,Iliesiu:2024cnh,DiUbaldo:2026rly,Maldacena:2026jqd}. The sum over topology is often framed as a maximalist `rule' for the gravitational path integral --- all topologies connecting all asymptotic boundaries must be included unless otherwise specified (see e.g. the philosophy laid out in \cite{Usatyuk:2024isz}). Within the context of holographic duality, this has led to interesting puzzles involving factorization of the dual Hilbert space or the partition function, as well as the dimension of the Hilbert spaces of closed universes (for which the dual system is presumably empty). It has been argued that certain bulk topologies must be included, while others must be canceled by UV contributions, such as `half-wormholes' \cite{Saad:2021rcu}, which are difficult to compute from first principles. Determining which topologies dominate or even contribute to the gravitational path integral asks for insight from a non-perturbative definition of quantum gravity.

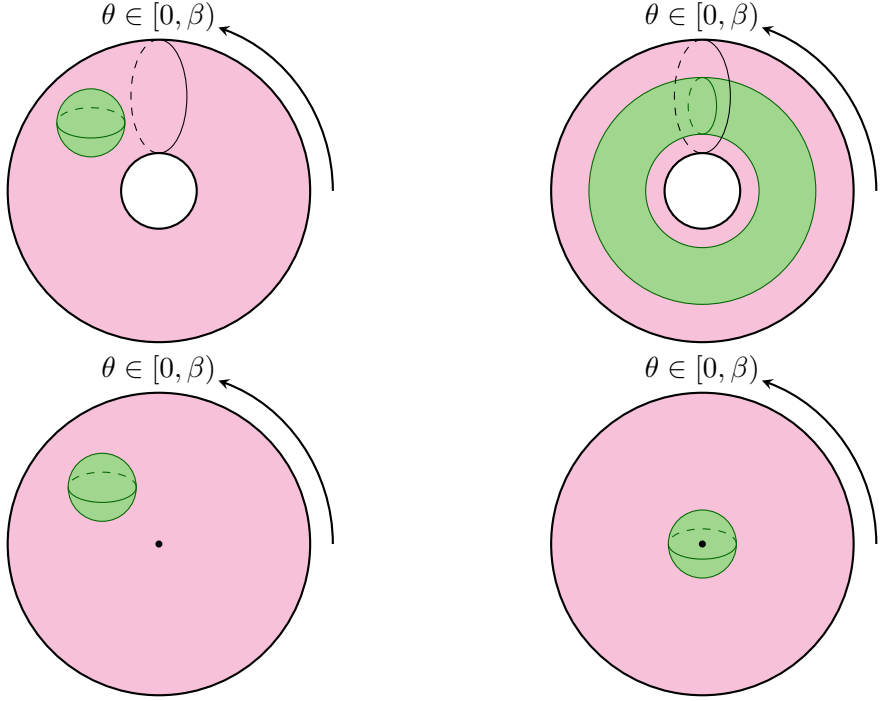
\begin{figure}[ht]
\centering
    \begin{subfigure}[b]{0.45\textwidth}
        \centering
        \begin{tikzpicture}
            \fill[magenta!30!white, even odd rule] (0,0) circle[radius=2cm] circle[radius=.5cm];
            \draw[thick] (0,0) circle (.5);
            \draw[thick] (0,0) circle (2);
            \draw (0,1.25) [partial ellipse = -90:90:.37 and .75];
            \draw[dashed] (0,1.25) [partial ellipse = 90:270:.37 and .75];

            \begin{scope}[xshift=-.9cm, yshift=.9cm]
                \fill[green, opacity=.35] (0,0) circle[radius=.45cm];
                \draw[black!60!green] (0,0) circle (.45);
                \draw[black!60!green] (0,0) [partial ellipse = 180:360:.45 and .2];
                \draw[black!60!green,dashed] (0,0) [partial ellipse = 0:180:.45 and .2];
            \end{scope}

            \draw[thick,-stealth] (2.3,0) arc[start angle = 0, end angle = 70, radius = 2.3];
            \node at (0,2.3) {$\theta\in[0,\beta)$};
        \end{tikzpicture}
    \end{subfigure}
    ~ 
    \begin{subfigure}[b]{0.45\textwidth}
        \centering
        \begin{tikzpicture}
            \fill[magenta!30!white, even odd rule] (0,0) circle[radius=2cm] circle[radius=.5cm];
            \draw[thick] (0,0) circle (.5);
            \draw[thick] (0,0) circle (2);

         \begin{scope}[xshift=0cm, yshift=0cm]
             \fill[green, opacity=.35, even odd rule] (0,0) circle[radius=1.5cm] circle[radius=.75cm];
                \draw[black!60!green] (0,0) circle (.75);
                \draw[black!60!green] (0,0) circle (1.5);
                \draw[black!60!green] (0,1.125) [partial ellipse = -90:90:.1875 and .375];
                \draw[black!60!green, dashed] (0,1.125) [partial ellipse = 90:270:.1875 and .375];
            \end{scope}

            \draw (0,1.25) [partial ellipse = -90:90:.37 and .75];
            \draw[dashed] (0,1.25) [partial ellipse = 90:270:.37 and .75];

            \draw[thick,-stealth] (2.3,0) arc[start angle = 0, end angle = 70, radius = 2.3];
            \node at (0,2.3) {$\theta\in[0,\beta)$};
        \end{tikzpicture}
    \end{subfigure}
    \\
    \begin{subfigure}[b]{0.45\textwidth}
        \centering
        \begin{tikzpicture}
            \fill[magenta!30!white, even odd rule] (0,0) circle[radius=2cm] circle[radius=0cm];
            \draw[thick] (0,0) circle (2);

            \begin{scope}[xshift=-.75cm, yshift=.75cm]
                \fill[green, opacity=.35] (0,0) circle[radius=.45cm];
                \draw[black!60!green] (0,0) circle (.45);
                \draw[black!60!green] (0,0) [partial ellipse = 180:360:.45 and .2];
                \draw[black!60!green,dashed] (0,0) [partial ellipse = 0:180:.45 and .2];
            \end{scope}

            \draw[thick,-stealth] (2.3,0) arc[start angle = 0, end angle = 70, radius = 2.3];
            \node at (0,2.3) {$\theta\in[0,\beta)$};
            \draw[very thick] (0,0) circle (.025);
        \end{tikzpicture}
    \end{subfigure}
    ~ 
    \begin{subfigure}[b]{0.45\textwidth}
        \centering

        \begin{tikzpicture}
            \fill[magenta!30!white, even odd rule] (0,0) circle[radius=2cm] circle[radius=0cm];
            \draw[thick] (0,0) circle (2);

            \begin{scope}[xshift=0, yshift=0]
                \fill[green, opacity=.35] (0,0) circle[radius=.45cm];
                \draw[black!60!green] (0,0) circle (.45);
                \draw[black!60!green] (0,0) [partial ellipse = 180:360:.45 and .2];
                \draw[black!60!green,dashed] (0,0) [partial ellipse = 0:180:.45 and .2];
            \end{scope}

            \draw[thick,-stealth] (2.3,0) arc[start angle = 0, end angle = 70, radius = 2.3];
            \node at (0,2.3) {$\theta\in[0,\beta)$};
            \draw[very thick] (0,0) circle (.025);
        \end{tikzpicture}
    \end{subfigure}
    \caption{Two relevant topologies to the Hawking-Page transition in the dual to a large-$N$ thermofield double state at inverse temperature $\beta$. \textbf{(Upper left)} In the thermal AdS saddle, the thermal circle is not contractible in the bulk and sphere worldsheets contribute linearly in $\beta$ to the partition function. These worldsheets do not contribute to the entanglement entropy of the dual thermofield double state. \textbf{(Upper right)} The leading contribution to the entropy in thermal AdS stems from worldsheets of torus topology, contributing $g_s^0\sim G_N^0$. \textbf{(Bottom left)} In the Euclidean black hole, the thermal circle is contractible and varying the semiclassical gravitation partition function results in a conical deficit at the horizon. Sphere worldsheets not intersecting this point do not contribute to the entropy for similar reasons to above. \textbf{(Bottom right)} Sphere topology worldsheets punctured by the conical deficit provide the leading contribution, $g_s^{-2}\sim G_N^{-1}$, to the black hole saddle; in the description of \cite{Susskind:1994sm,Ahmadain:2022eso} these worldsheets require an off-shell description of string theory.}
    \label{fig:Hawking-Page-Worldsheet}
\end{figure}

In this paper we will take some initial steps in this direction, focusing on simple examples of exchange of dominance in the Euclidean computation of entanglement entropy. To be concrete, we explore the Hawking-Page transition in the entanglement of the thermofield double state and the phenomena of replica wormholes which reproduce the Page curve of black hole evaporation. As we will explain below, both examples can be phrased in terms of the (non)contractibility of an $S^1$ cycle in the bulk description. Our probe of choice for this topology change is the string worldsheet, and the point of view we adopt in this paper is to take for granted the proposal that `t Hooft ribbon diagrams of matrix theories \cite{tHooft:1973alw} are the non-perturbative realization of these worldsheets. This is itself a form of holographic duality in which a geometric bulk description -- namely the background in which the worldsheets propagate -- emerges from the dynamics of a large-$N$ quantum system. This intuition has been central in many developments in the study of matrix models (e.g.\cite{Brezin:1977sv,Witten:1979kh,Ginsparg:1993is,Klebanov:1991qa,Saad:2019lba} as a far from exhaustive set) and in recent years has been made extremely precise \cite{Gopakumar:2003ns,Gopakumar:2004qb,Gopakumar:2005fx,Aharony:2006th,Gopakumar:2022djw,Gopakumar:2024jfq,Gaiotto:2024dwr,Gaiotto:2025hjn}. Our work is not yet at the same level of rigor as these analyses, we will instead take them as evidence for utilizing the `t Hooft worldsheet as a non-perturbative proxy for the string worldsheet.\\\\
\emph{The first result of this paper is that the 't Hooft diagrammatics of large-$N$ gauged}\footnote{Gauging in 0+1d is synonymous with projection onto the singlet sector. The dynamical projection onto this singlet sector that is suggested in \cite{Maldacena:2018vsr} to occur in holographic systems suffices.}\emph{ matrix quantum mechanics (MQM) can systematically reproduce the exchange of dominance in bulk topologies contributing to a R\'enyi entropy.} This includes the well-known Hawking-Page transition, as well as the exchange from the na\"ive disconnected bulk topology (the ``Hawking saddle'') to a replica wormhole topology. We emphasize that this analysis is obtained within a single large-$N$ theory without recourse to ensemble averaging, as in JT gravity or averaged CFTs \cite{Penington:2019kki,Almheiri:2019qdq,Geng:2025efs}.\\
\\
Both the definition and computation of entanglement entropy in the target space of string theory is plagued with subtleties. Even barring questions of factorization, it is expected that the area law contribution, $A/4G_N$, requires non-perturbative worldsheet physics over which we do not have complete control.\footnote{See \cite{Ahmadain:2022eso,Ahmadain:2025pox} for recent reviews and progress; see also \cite{HalderJafferisWorldsheet2023} for a sphere-level computation of BTZ entropy.} Instead we will take inspiration from the replica trick computations of Susskind and Uglum \cite{Susskind:1994sm}. In that work the classical black hole entropy, $A/4G_N$, is reproduced from sphere topology string worldsheets that are punctured by a conical deficit running along the fixed point of replica symmetry, in this case the black hole horizon. We will use this simple fact as an order parameter for topology change: namely, the (in)ability for genus zero string worldsheets to be punctured by a bulk conical deficit is directly determined by the bulk topology, as illustrated in Figure \ref{fig:Hawking-Page-Worldsheet}. 

On the matrix side we will compute R\'enyi entropies directly in 0+1 dimensional MQM. The Hilbert space is manifestly well defined, as the space of $L^2$-normalizeable functions of bosonic matrices potentially tensored with finite-dimensional factors for any fermionic matrices. This framework also allows us to prepare wavefunctions and compute entanglement without committing ourselves to a specific Hamiltonian. Given some collection of matrix variables $X^i$, we focus on `single-trace' wavefunctions of the form
\begin{equation}\label{eqn:intro-schematic-wavefunction}
\Psi(X^i) = \exp(-\Tr\mc S(X^i)),
\end{equation}
where $\mc S$ is an $O(N^0)$-degree polynomial in $X^i$ with a good `t Hooft limit (meaning that the coefficients of traces with $k$ $X^i$ insertions scale as $N^{(2 - k)/2}$). The purpose of such an ansatz is that correlation functions of single-trace operators $\langle \Tr (X^{i_1}X^{i_2}\ldots) \ldots \Tr (X^{j_1}X^{j_2}\ldots) \rangle$ generate ribbon diagrams just as any `t Hooft expansion without reference to dynamics: the role played by a Hamiltonian or Lagrangian is replaced by the weight $\Tr\mc S(X^i)$. Furthermore, as we show in detail in \S\ref{ssec:wavefunctions}, the $n^\text{th}$ R\'enyi entropy of some matrix subsystem of \eqref{eqn:intro-schematic-wavefunction} is computed by a matrix integral that itself generates an `t Hooft expansion, and so in the perspective described above, manifestly defines a string background.

The signature of topology change from this perspective is the leading large-$N$ behavior of the `t Hooft diagrams contributing to the R\'enyi entropy of such wavefunctions. In this computation, there is no conical defect; the above transition arises from a deconfinement transition in the matrix integrals wrapping the replica circle. The relationship between the bulk Hawking-Page transition and boundary deconfinement transition is already well-known, as is the role of deconfinement in the large-$N$ scaling of `t Hooft diagrams \cite{Atick:1988si,FuruuchiThermal2005,FuruuchiLectures2006}. Here we will present concrete wavefunctions exemplifying this physics. 

We stress that we do \textit{not} prove that our ans\"atze are in fact black hole duals, although they are inspired by schematic proposals \cite{Banks:1997hz,Berenstein:2002jq,Berkowitz:2016znt,Gautam:2022akq,Martinec:2026wuu}. Regardless, the diagrammatic expansion that they generate can, under the state and saddle assumptions below, diagnose bulk topology change -- the thermal Hawking-Page transition, and the transition from the Hawking saddle to the replica wormholes saddle. Notably, in our work this transition does \textit{not} occur for general MQM states with a flat entanglement spectrum across some microcanonical window --- or to restate it in the language of \cite{Guo:2021blh}, we are able to distinguish between black holes and other energetic bound states of branes. We interpret this result as a schematic for a non-perturbative definition of string theory in replica wormhole backgrounds.

\begin{figure}[ht]
\centering
\includegraphics[width=0.7\textwidth]{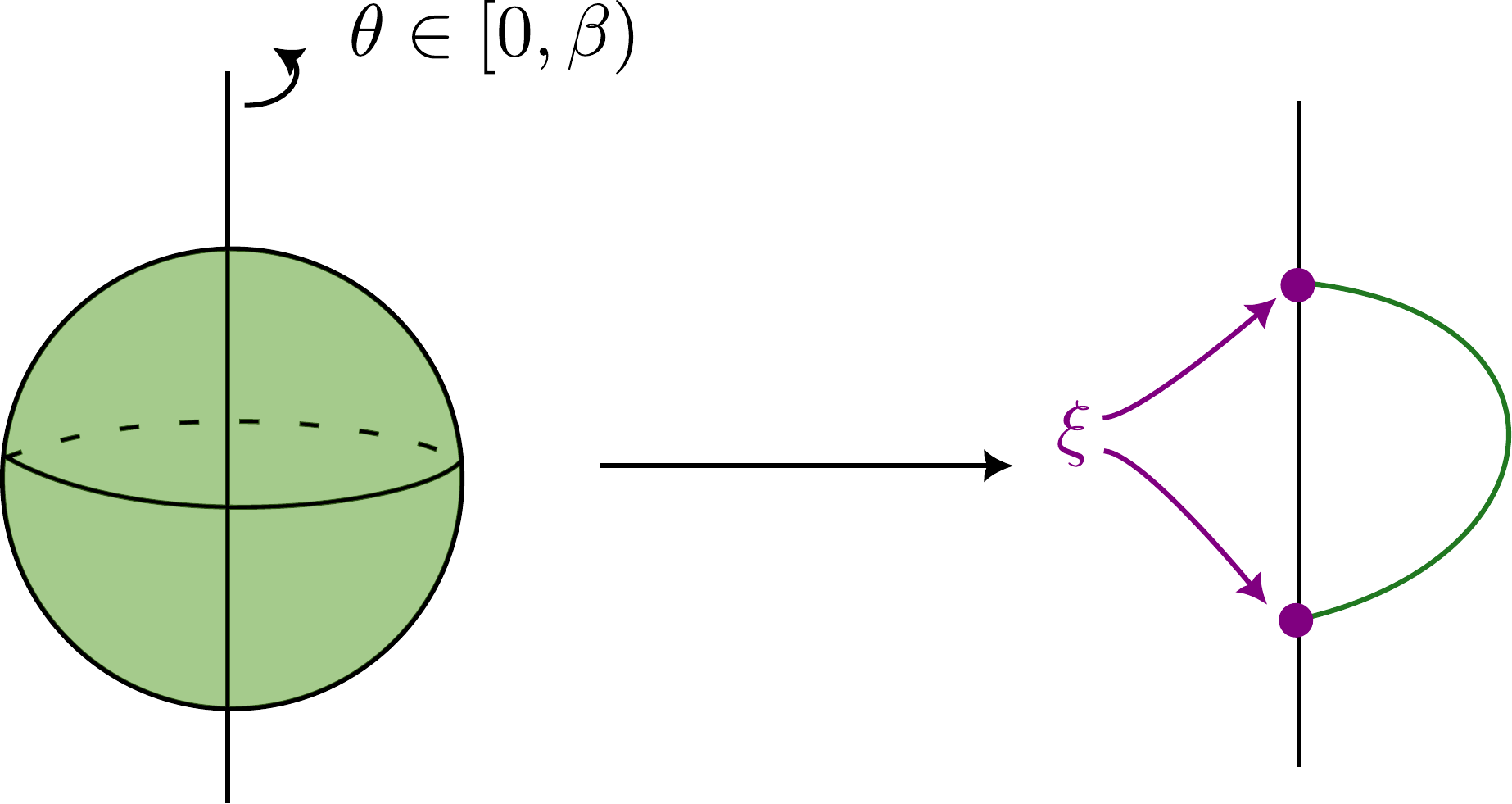}
\caption{The edge mode physics proposed in \cite{Susskind:1994sm}. If the area-law entanglement across horizons in gravity is due to punctured closed strings, (as depicted on the left), then a time-slice attributes this entanglement to open string edge modes, $\xi$ (as depicted on the right).
}\label{fig:SU-open-strings}
\end{figure}

We will press this story further to elucidate a more fine-grained picture of bulk entanglement. In the Susskind-Uglum story, an angular slicing along the replica circle highly suggests that the area law entropy arises from the entanglement of open strings attached to the horizon, as depicted in Figure \ref{fig:SU-open-strings}. Thus Hilbert space of closed strings should somehow be extended to include these open string sectors which are the string-theoretic instantiation of `entanglement edge modes':\footnote{See \cite{Donnelly:2016auv} for a review of the role of edge modes in gauge theory and gravity.}
\begin{equation}\label{eqn:SU-Hilbert-space-factorization}
    \mathcal{H}_{\text{closed}} \subset \mathcal{H}_{\text{open}}^\text{in} \otimes \mathcal{H}_{\text{open}}^\text{out}.
\end{equation}
This decomposition is also deeply tied to the bulk topology: closed string diagrams in the thermal AdS saddle see no conical deficit and the Lorentzian continuation contains no bifurcate horizon on which to anchor open string edge modes. In this case the Hilbert space naturally decomposes as $\mathcal{H}_{\text{closed}} = \mathcal{H}^\text{in}_{\text{closed}} \otimes \mathcal{H}^\text{out}_{\text{closed}}$. Here we look for a non-perturbative description of this decomposition at the level of the worldsheet.\\
\\
\emph{As a second result, we will find a natural decomposition of the `t Hooft diagrams admitting degrees of freedom interpretable as open-string edge modes.} As required by basic consistency with the role target space topology plays in the decomposition \eqref{eqn:SU-Hilbert-space-factorization} \cite{Susskind:1994sm,Lewkowycz:2013nqa}, we will see that our local angular expansion requires localization beyond deconfinement, where the replica circle is contractible in the proposed dual bulk description. Thus we illustrate a toy mechanism for how bulk topology change can lead to the Hilbert space factorization \eqref{eqn:SU-Hilbert-space-factorization} in a non-perturbative definition of string theory.

\subsection{Relationship to Past Work}

The search for a large-$N$ description of entanglement edge modes in string theory is not new. Donnelly and Wong gave an explicit closed/open Hilbert-space construction in the Gross--Taylor string \cite{DonnellyWongEBranes2017}. The authors of \cite{Donnelly:2020teo,Jiang:2020cqo} made the program precise in topological string theory.\footnote{The `t Hooft limit already enters \cite{Jiang:2020cqo}; see also \cite{Wong:2026ftn}. Our focus here is the replica integrals of singlet MQM wavefunctions. It would be interesting to understand how this work relates to theirs.}
A line of papers has specifically studied target space entanglement (and related ideas) in matrix models \cite{Hartnoll:2015fca,Das:2022nxo,Han:2019wue,Das:2020jhy,Hampapura:2020hfg,Gautam:2022akq,Frenkel:2021yql,Frenkel:2023aft,Fliss:2025kzi,Fliss:2025omb}.
The current work focuses explicitly on the `t Hooft limit of MQM entanglement calculations in order to derive insights towards string theory.

Our description of how punctured sphere worldsheets arise from ribbon diagrams in an entropy calculation is, in some sense, a modern re-interpretation of older results. In the $c=1$ matrix model the non-singlet sector of MQM generates `t Hooft worldsheets with winding modes \cite{Klebanov:1991qa,Kazakov:2000pm}. A key difference in our framework is we do \textit{not} pass to a non-singlet sector --- we treat the adjoint $U(N)$ symmetry as an honest gauge symmetry, allowing our results to be generalized to higher dimensional models. Furthermore, in \cite{Kazakov:2000pm} the proliferation of worldsheet punctures is identified as an infinite-order worldsheet BKT transition corresponding to a bulk Hagedorn phase, whereas for us in this work it is tied to the deconfinement transition. This relationship between the deconfinement transition and string worldsheet has also been known since Atick and Witten \cite{Atick:1988si} and its role in the proliferation of vortices on the worldsheet was developed in \cite{FuruuchiThermal2005,Furuuchi:2006st,FuruuchiLectures2006}, including the relation to the contractibility of the thermal circle. Here we go beyond this traditional picture by providing concrete single-trace ans\"atze which not only make this physics manifest, but also provide the natural framework for connecting it to the appearance of open string edge modes.

Our setup for studying the replica wormhole transition (in \S\ref{sec:replica-wormholes}) is essentially that of Liu and Vardhan \cite{Liu:2020jsv} --- we assume that at late times the reduced density matrix on the relevant MQM degrees of freedom is perturbatively close to a microcanonical ensemble. Relatedly, the Haar averages discussed in \cite{Liu:2020jsv} are over unitaries acting on the full Hilbert space or a microcanonical energy window, whereas ours are over color-space gauge transformations implementing singlet projection. Their general equilibrium approximation also applies to fixed-Hamiltonian evolution without ensemble averaging. The key difference in our analysis is that the bulk topology emerges in an entirely different manner --- our planar diagrams are worldsheets living in the replica wormhole geometry, and are not the replica wormhole geometry themselves. The core lesson of this paper is that we must deduce the bulk topology by reading off NLSM like data from the `t Hooft worldsheet: in our analysis \textit{not every} maximally entangled chaotic system produces diagrams that predict a replica wormhole-like topology in a putative bulk dual --- the relationship of the microcanonical energy to a deconfinement threshold is crucial.

\subsection{Organization of Paper}

The results of this paper are summarized as follows. In \textbf{Section \ref{ssec:wavefunctions}} we establish the ansatz wavefunctions to be used throughout the paper; this serves partially as a vehicle for establishing notational conventions. In \textbf{Section \ref{sec:contractible-cycles}} we show how entanglement computations in these single-trace wavefunctions give rise to a non-perturbative definition of the punctured sphere worldsheets of \cite{Susskind:1994sm}.
The thermofield double-like example developed in Appendix \ref{app:BMN-deconfinement} provides a proof of concept of this transition within our class of ans\"atze. In \textbf{Section \ref{sec:thermofield-double}} we address how this transition leads to the appearance of open string edge modes anchored at a bifurcate horizon. In \textbf{Section \ref{sec:replica-wormholes}} we apply our technology to the case of replica wormholes and show that the `t Hooft diagrammatics capture the transition from the Hawking saddle to the replica wormhole saddle. We conclude the paper in \textbf{Section} \ref{sec:discussion} with a discussion of the results as well as remaining questions and future open directions.

\section{`t Hooft-like Wavefunction Ans\"atze}\label{ssec:wavefunctions}

In this brief section we describe our ans\"atze of interest for MQM wavefunctions. We will first introduce our notational conventions for the zoo of objects and indices that arise in MQM, matrix chains, and quiver theories.

\subsection{Notational Conventions}

The large-$N$ theories we are interested in have degrees of freedom defined on auxiliary color spaces. We always take these spaces to be $\mathbb{C}^N$, for some $N$, unless otherwise specified. We take $a,b,c,\ldots$ to be indices on color spaces. Matrix chains and quiver theories associate a color space to each node in some graph. We label these nodes by sans-serif letters, $\ms k$, $\ms l,\ldots$, so the associated color space has complex dimension $N_{\ms k}$. Traces over color spaces are always capitalized, such as in the expression $\Tr[X^2]$. Each color space supports the action of $U(N_{\ms k})$, which we refer to as `gauge transformations' regardless of whether we wish to consider them as gauge or global symmetries.

The degrees of freedom we are interested in are classified by how they transform under the full gauge group 
\begin{equation}\label{eqn:gauge-group}
G = \prod_{\ms k} G_{\ms k} = \prod_{\ms k}U(N_{\ms k}).
\end{equation}
A set of $D$ $N_{\ms k} \times N_{\ms k}$ matrices transforming in the adjoint representation under $U(N_{\ms k})$ is denoted $X_{\ms k}^i$. The superscript $i$ runs from $1$ to $D$, and the subscript ${\ms k}$ labels which node the matrix belongs to. We typically assume adjoint degrees of freedom are Hermitian matrices, and make it clear from context when this is not the case. Fundamental representations of $U(N_{\ms k})$ are denoted by lowercase Greek letters such as $\phi^i_{\ms k}$, $\eta^i_{\ms k}$, or $\xi^i_{\ms k}$.

We will treat the above degrees of freedom quantum mechanically. For degrees of freedom $X^i_{\ms k}$ we define an associated Hilbert space $\mathcal{H}_{\ms k}$ of square-normalizeable functions of the matrices $X^i_{\ms k}$. Traces over Hilbert spaces will be lowercase, as in $\tr[\rho^n]$. In evaluating entanglement entropies through replica trick, replica indices are denoted using $q,r, \ldots$.

Expectation values are computed through matrix integrals weighted by wavefunctions. For notational simplicity we will denote all integration measures simply as
\beq
    [\dd X^i]~,\qquad [\dd U]~,\qquad [\dd \phi^i]~,\qquad\text{etc.}
\eeq
For unitary matrices this shorthand denotes the Haar invariant integration measure, while for Hermitian matrices and fundamental degrees of freedom this is the flat integration measure over their $\mathbb C$-valued entries. When this distinction is not clear from context we will state so explicitly.

\subsection{Single-trace and Gauge Invariant Wavefunctions}

The wavefunctions we consider in this paper are exponentials of single-trace expressions. For example, the simplest class of wavefunctions we might consider are single-trace exponentials of a set of $N \times N$ matrices $X$:
\begin{equation}\label{eqn:ansatz-1}
\Psi(X^i) = \mathcal{\sqrt{N}}\exp\left(-\Tr\mc S(X^i)\right),
\end{equation}
for some complex polynomial of the matrices $\mc S(X^i)$. We sometimes refer to $\mc S$ as an `action' or a `weight' even though it a priori has nothing to do with an action principle associated to a Lagrangian. $\mathcal{N}$ is the appropriate normalization. The simplest expectation values are computed as
\begin{equation}\label{eqn:expectation-values-1}
\langle \Tr[X^p] \rangle = \mathcal{N}\int \prod_i[\dd X^i]\, \exp(-\Tr[\mc S(X^i)+\bar{\mc S}(X^i)]) \Tr[X^p].
\end{equation}
The \textit{raison d'\^etre} of the ansatz \eqref{eqn:ansatz-1} is that such expectation values (or the connected correlators of multi-trace insertions) are the generating functions of `t Hooft expansions in the appropriate scaling of the coefficients in the polynomial $\mc S(X^i)$.

As stated the ansatz \eqref{eqn:ansatz-1} is extremely restrictive. A generic $U(N)$ invariant function of the $X^i$ should allow for all possible multi-trace polynomials. This being said, the expectation values \eqref{eqn:expectation-values-1} generates are already interpretable as coming from noncritical $c < 1$ string theories in the `t Hooft limit \cite{Ginsparg:1993is}. We may vastly generalize this ansatz by integrating in auxiliary variables. As an instructive example consider an additional set of $N \times N$ matrices $\tilde{X}^i$ transforming in the adjoint representation of $U(N)$. We can build the family of wavefunctions
\begin{equation}\label{eqn:Psi-kappa}
\Psi_\kappa(X^i) = \sqrt{\mathcal{N}} \int \prod_i[\dd\tilde{X}^i]\, \exp(-\Tr[\mc S(X^i) + \tilde{\mc S}(\tilde{X}^i) + \kappa \sum_i X^i \tilde{X}^i X^i \tilde{X}^i])
\end{equation}
where $\kappa$ is a non-dynamical parameter. $\Psi_\kappa(X^i)$ is still invariant under the gauge transformation $X^i \rightarrow U X^i U^{\dag}$ for all $\kappa$. Integrating out $\tilde X^i$ yields an effective multi-trace wavefunction of $X^i$. For instance, if $\tilde{\mc S}(\tilde X^i) = \sum_i (\tilde X^i)^2 $ then the $\tilde{X}^i$ integral is purely Gaussian resulting in an effective wavefunction 
\begin{equation}\label{eqn:new-Psi-kappa}
\begin{split}
\Psi_\kappa(X^i) &= \sqrt{\mathcal{N'}} \exp(-\Tr\mc S(X^i)) \prod_i \det(\mathbb{1}_{N^2} + \kappa X^i \otimes X^i)^{-1/2}\\
&= \sqrt{\mathcal{N'}} \exp(-\Tr\mc S(X^i) - \frac{1}{2}\sum_i\sum_{n=1}^{\infty} \frac{(-1)^{n + 1}\kappa^n}{n}(\Tr[(X^i)^n])^2).
\end{split}
\end{equation}
We have absorbed additional constant factors into the  new normalization $\mathcal{N'}$. We can see that integrating out $\tilde{X}^i$ has left a new gauge-invariant wavefunction with an infinite series of multi-trace terms in the exponent. It should be clear that more general choices of single-trace $\Tr\tilde{\mc S}(\tilde{X}^i)$ and interaction terms in \eqref{eqn:Psi-kappa}, as well as more auxiliary variables, can build arbitrarily complex wavefunctions.

A similar strategy may be used for building entangled gauge-invariant wavefunctions for multi-boundary wavefunctions. Let us consider a two-site MQM system with labels $L$ and $R$, and degrees of freedom $X_L^i$ and $X_R^i$. We refer to the unitary transformations acting on $X_L^i$ as $G_L$ and the unitary group acting on $X_R^i$ as $G_R$. This is a thermofield double-like setup that we analyze in \S\ref{sec:thermofield-double}. We wish for these wavefunctions to be invariant under the full gauge group \eqref{eqn:gauge-group}, so that they satisfy
\begin{equation}
\Psi(X_L^i,X_R^i) = \Psi(U X_L^i U^{\dag}, X_R^i) = \Psi(X_L^i,UX_R^iU^{\dag}), \quad \forall \;U \in U(N).
\end{equation}
A single-trace action that is invariant under both $G_L$ and $G_R$ can be built with the help of an auxiliary unitary $U \in U(N)$:\footnote{We may alternatively use bifundamental degrees of freedom $K^i$:
\begin{equation}
\Psi_\kappa(X_L^i,X_R^i) = \sqrt{\mathcal{N}}\int \prod_i[\dd K^i]\, \exp(-\Tr[\mc S_L(X^i_L)+\mc S_R(X^i_R) + \kappa \sum_{i}X^i_L K^i X_R^i {K^i}^{\dag}])~,
\end{equation}
although we will not make use of this in this paper.}
\begin{equation}\label{eq:EntStateExamp}
\begin{split}
\Psi_\kappa(X_L^i,X_R^i) = \sqrt{\mathcal{N}}\int [\dd U]\, \exp\left(-\Tr[\mc S_L(X^i_L)+ \mc S_R(X^i_R)-\kappa\sum_iX^i_L U X_R^i U^{\dag}]\right)~.
\end{split}
\end{equation}
Gauge invariant wavefunctions of this form will play a central role throughout this paper.\footnote{For a single matrix pair, the unitary integral appearing here is precisely of the Harish-Chandra-Itzykson-Zuber (HCIZ) type \cite{harish1957differential,Itzykson:1979fi} which is exactly solvable, and whose large-$N$ behavior scales as $N^2$ \cite{Matytsin:1993iq,Bun:2014dha}. Ref. \cite{Matytsin:1993iq} in particular highlights the connection between the large-$N$ expansion of HCIZ integral and the hydrodynamics of the $X^i_{L/R}$ eigenvalues. We thank Wayne W. Weng for discussions on this point.}

\subsection{Example: The BMN thermofield double}\label{ssec:BMN-TFD-examp}
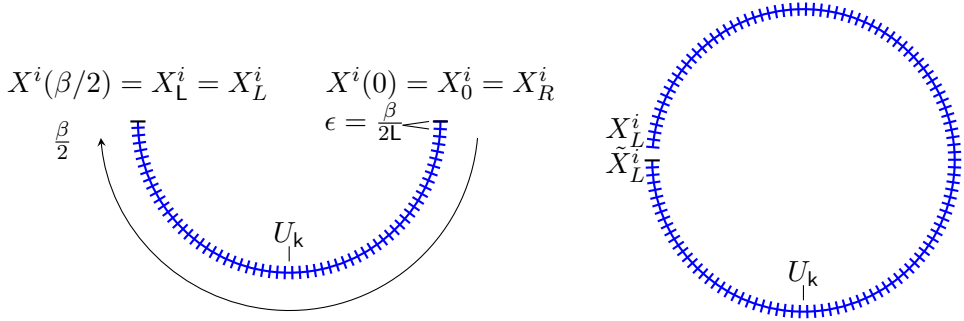
\begin{figure}[ht]
    \centering
    \begin{subfigure}[b]{0.45\textwidth}
        \centering
        \begin{tikzpicture}
            \draw[thick,blue,postaction = {draw, decorate,
            decoration = {ticks, segment length=1mm}}]
            (0,0) [partial ellipse = 0:-180:2 and 2];
            \draw[-{stealth}] (0,0) [partial ellipse = -5:-175:2.5 and 2.5];
            \draw[thick] (-2.1,0) -- (-1.9,0);
            \draw[thick] (2.1,0) -- (1.9,0);
            \node at (-3,-.25) {$\frac{\beta}{2}$};
            \node at (0,-1.5) {$U_{\mathsf{k}}$};
            \draw[thin] (0,-1.65) -- (0,-1.85);
            \draw[thin] (1.85,0) -- (1.5,-.5mm);
            \draw[thin] (1.85,-1mm) -- (1.5,-.5mm);
            \node at (1,0) {$\epsilon=\frac{\beta}{2\mathsf{L}}$};
            \node at (2,.5) {$X^i(0)=X^i_0=X^i_R$};
            \node at (-2,.5) {$X^i(\beta/2)=X^i_{\mathsf{L}}=X^i_L$};
        \end{tikzpicture}
    \end{subfigure}
    \begin{subfigure}[b]{0.45\textwidth}
        \centering
        \begin{tikzpicture}
            \draw[thick,blue,postaction = {draw, decorate,
            decoration = {ticks, segment length=1mm}}]
            (0,0) [partial ellipse = 175:-180:2    and 2];
            \draw[thick] (-2.1,0) -- (-1.9,0);
            \node at (0,-1.5) {$U_{\mathsf{k}}$};
            \draw[thin] (0,-1.65) -- (0,-1.85);
            \node at (-2.35,-.05) {$\tilde X^i_L$};
            \node at (-2.35,.4) {$X^i_L$};
        \end{tikzpicture}
    \end{subfigure}
\caption{\textbf{(Left)} Discretizing Euclidean time into $\ms L$ time-steps leads to a lattice gauge theory description of the thermofield double state, with matrix degrees of freedom, $X^i_{\ms k}$, living on the nodes and unitaries, $U_{\ms k}$, living on the links. \textbf{(Right)} The reduced density matrix as a lattice gauge theory on a chain of length of $2\ms L$.}\label{fig:discPI}
\end{figure}

To ground the above discussion, we study the following wavefunction, which is a lattice gauge theory realization of the thermofield double of the matrix model of Berenstein, Nastase, and Maldacena (BMN) \cite{Berenstein:2002jq}. We will focus on the bosonic sector of this model although the fermions can be included with minor modification, as we do in Appendix \ref{app:BMN-deconfinement}:
\begin{equation}\label{eq:TFDasquiver}
\begin{split}
\Psi^\text{BMN}_{\ms L}(X^i_L,;X^i_R,) = \sqrt{\mc N_{\ms L}}\int &\prod_{\ms k=1}^{\ms L-1}\prod_i [\dd X_{\ms k}^i] \prod_{\ms k=1}^{\ms L} [\dd U_{\ms k}]\\
&\times\exp\left(-\frac{2N\ms L}{\beta\,g^2}\sum_{\ms k=1}^{\ms L}\sum_{i=1}^9 \Tr[(U_{\ms k} X^i_{\ms k - 1}U_{\ms k}^{\dag} - X^i_{\ms k})^2]\right)\\
&\qquad\times\exp\left(- \sum_{\ms k=0}^{\ms L}w_{\ms k}\frac{N\beta}{2g^2\ms L}\Tr V_{\bmn}(X^i_{\ms k})\right)~,
\end{split}
\end{equation}
where
with weights $w_0=w_{\ms L}=\tfrac12$ and $w_{\ms k}=1$ for $1\leq\ms k\leq\ms L-1$, and bosonic potential 
\begin{equation}\begin{split}\label{eq:VBMN}
    V_{\bmn}(X^i) = &-\frac{1}{2}\sum_{i\neq j}[X^i,X^j]^2+i\frac{2\mu}{3}\sum_{i,j,k=1}^3\epsilon_{ijk}X^iX^jX^k\\
    &+\left(\frac{\mu}{3}\right)^2\sum_{i=1}^3(X^i)^2+\left(\frac{\mu}{6}\right)^2\sum_{i=4}^9(X^i)^2~.
\end{split}
\end{equation}
Our example wavefunction, \eqref{eq:TFDasquiver}, is related to the standard thermofield double state in the following way. The thermofield double can be prepared by the Euclidean path integration over the half-circle of Euclidean time, $\beta/2$, with boundary conditions $X^i(0)=X^i_R$, and $X^i(\beta/2)=X^i_L$.  
We discretize this path-integral along the Euclidean time direction into a total of $\ms L$ units of $\epsilon=\frac{\beta}{2\ms L}$, indexed by a `quiver' index $\ms k$. See Figure \ref{fig:discPI} for a cartoon. The discretized parallel transport operator
\beq\label{eq:paratransOp}
    U_{\ms k}=Pe^{i\epsilon A_{\ms k}}~.
\eeq
gauges the $U(N)$ symmetry in Euclidean time and leads directly to \eqref{eq:TFDasquiver}. We refer an interested reader to App. \ref{app:BMN-deconfinement} for details.

\begin{figure}[ht]
    \centering
    \begin{tikzpicture}
    \draw[very thick,blue]
    (0,0) [partial ellipse = 0:-180:2 and 2];
    \draw[very thick] (-2.1,0) -- (-1.9,0);
    \draw[very thick] (2.1,0) -- (1.9,0);
    \node at (0,-1.5) {$U$};
    \node at (2,.5) {$X^i_R$};
    \node at (-2,.5) {$X^i_L$};
    \end{tikzpicture}
\caption{An extreme discretization of the wavefunction depicted on the left of Fig. \ref{fig:discPI}, consisting of a single time step.}\label{fig:L1discWF}
\end{figure}
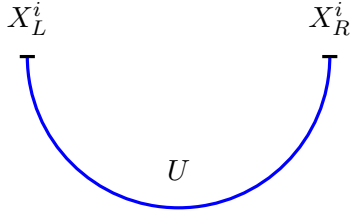

We will ultimately be interested in the entanglement entropy of the reduced density matrix associated to tracing out the $R$ sector of \eqref{eq:TFDasquiver}, as depicted on the right of Fig. \ref{fig:discPI}:
\beq
    \rho^\text{BMN}_L[X_{L},\tilde X_{L}]=\int[\dd X_R][\dd U]\Psi^\text{BMN}_{\ms L}(X_{L};X_R)^\ast\Psi^\text{BMN}_{\ms L}(\tilde X_{L};UX_RU^\dagger)~.
\eeq
The path-integral description of the R\'enyi entropy can be computed by a lattice gauge theory on a closed chain of length $2n\ms L$. This path integral displays much of the salient physics that we will explore below. Namely, as we show in detail in Appendix \ref{app:BMN-deconfinement}, the R\'enyi path integral displays a deconfinement transition in the parallel transport operator, \eqref{eq:paratransOp}. One important upshot of that analysis is the robustness of this transition to the discretization of Euclidean time. Indeed, this transition exists even in the most violent discretization of Euclidean time with $\ms L=1$! That is to say, while a mild departure from the continuum thermofield double, all of the important physics is captured by the simpler entangled wavefunction 
\beq\label{eq:TFDbosonicsimplified}
    \Psi_1^\text{BMN}(X_L^i,X_R^i)=\sqrt{\mc N_1}e^{-\Tr\mc S(X_L)-\Tr\mc S(X_R)}\int[\dd U]\exp\left(\frac{4N}{\beta g^2}\sum_i\Tr[X^i_LUX^i_RU^\dagger]\right)~,
\eeq
as depicted in Figure \ref{fig:L1discWF}. Here $\mc S(X)=\frac{2N}{\beta g^2}\sum_i(X^i)^2+\frac{N\beta}{4g^2}V_{\bmn}(X^i,0)$. We invite the reader to note that \eqref{eq:TFDbosonicsimplified} is precisely of the form considered above, \eqref{eq:EntStateExamp} where $\kappa=\frac{4N}{\beta g^2}$. Thus the reader is encouraged to keep \eqref{eq:TFDbosonicsimplified} as a concrete example wavefunction whose `t Hooft expansion diagnoses topology change in the R\'enyi entropies of a (putative) bulk string theory.

\section{Bulk Topology from the `t Hooft Expansion: Contractible vs. Non-Contractible $S^1$ cycles}\label{sec:contractible-cycles}

In this section we ask how the planar expansion reproduces the Susskind-Uglum picture of closed strings punctured by a conical deficit. In particular we look for the mechanism by which the `t Hooft worldsheets capture the transition from thermal AdS to the Euclidean black hole, as a direct computation of an entanglement entropy. To be specific, we will show that our wavefunction ans\"atze probe a simple aspect of the bulk topology, namely whether or not a certain boundary $S^1$ cycle (the cycle that winds around replica number) is contractible in the dominant bulk saddle point geometry. We call this cycle the `replica circle,' as a nomenclatural generalization of `thermal circle.' Our arguments will apply to a large class of single trace matrix wavefunctions, and we will present them as such, however also include the specific lattice thermofield double of the BMN model, \eqref{eq:TFDbosonicsimplified}, described in the previous section.

\begin{figure}[ht]
\centering
\includegraphics[width=0.4\textwidth]{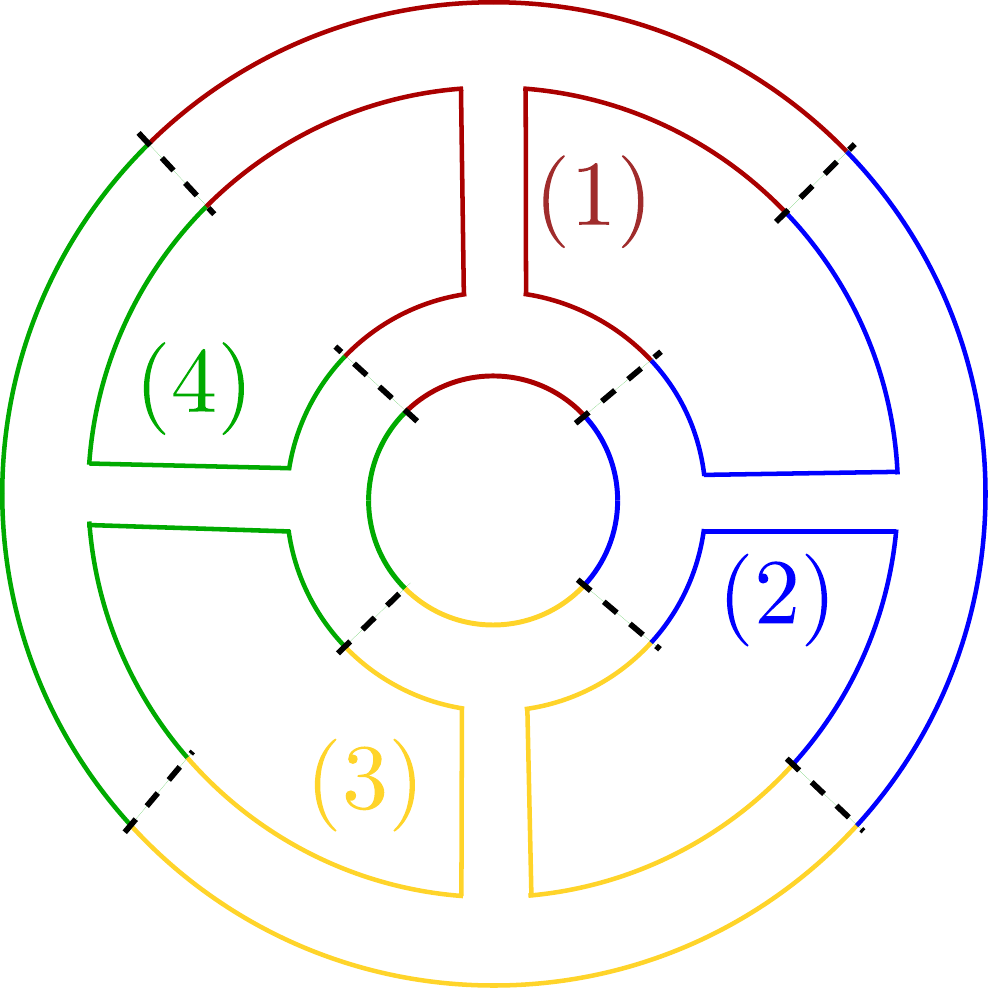}
\caption{A possible sphere-topology `t Hooft diagram generated by the partition function \eqref{eqn:nth-Renyi-integral} with a cubic interaction, $\mc S\supset X^3$, for the $n=4$ R\'enyi entropy. Propagators and vertices are labeled by $(q)$ running from $(1)$ to $(4)$, which we also distinguish by color for visual aid. This diagram wraps the replica circle --- there are two propagators (the innermost circle and the outermost circle) that cycle through all the labels $(1)$ through $(4)$ in a sequence with nontrivial winding number.}\label{fig:colored-ribbons-bare}
\end{figure}

We begin with the concrete  gauged single-trace ansatz, \eqref{eq:EntStateExamp}; a more general discussion of gauged single-trace wavefunctions can be found in App. \ref{app:deconfinement-topology}. We imagine computing the entanglement entropy between $X_L^i$ and $X_R^i$. The reduced density matrix on the $X_L^i$ degrees of freedom is 
\begin{equation}
\begin{split}
    \rho(X_L^i,\tilde{X}_L^i) = \mathcal{N} \int \prod_i[\dd X_R^i][\dd U][\dd V] &\exp(-\Tr[\mc S_L(X_L^i)+\bar{\mc S}_L(\tilde{X}_L^i)+2\Re \mc S_R(X_R^i)])\\
    \times&\exp(\kappa\sum_i\Tr[X_L^iVX_R^iV^\dagger+\tilde{X}_L^iU X_R^iU^\dagger])~,
\end{split}
\end{equation}
and the entanglement entropy is given as the von Neumann entropy of this reduced density matrix which we compute through the replica trick: 
\begin{equation}\label{eqn:vNfromSn}
    S_\text{vN}[\rho]=\lim_{n\rightarrow1}S_n[\rho]
\end{equation}
where $S_n[\rho]$ is the $n^\text{th}$ R\'enyi entropy of $\rho$:
\beq
    S_n[\rho]=\frac{1}{1-n}\log\frac{Z_n}{(Z_1)^n}~,\qquad\qquad Z_n:=\tr[\rho^n]~.
\eeq
Expanding about $n=1$, we can alternatively express $S_\text{vN}$ as
\begin{equation}\label{eqn:von-Neumann-replica}
    S_\text{vN}[\rho]=-\lim_{n\rightarrow 1}\pa_n\left(\log Z_n-n \log Z_1\right)
\end{equation}
Both expressions, \eqref{eqn:vNfromSn} and \eqref{eqn:von-Neumann-replica}, require us to analytically continue $n$, a point that we will return to below. To compute $Z_n$ we introduce a replica index $q$ as $X_{L,R;q}^{i}$ identifying $q$ with $q + n$, writing

\begin{equation}\label{eqn:nth-Renyi-integral}
\begin{split}
    Z_n = \mathcal{N}^n \int \prod_{q=1}^n\prod_{i} [\dd X_{L;q}^{i}][\dd X_{R;q}^{i}]&[\dd U_q][\dd V_q] \exp(- \sum_{q=1}^n\Tr[2\Re \mc S_L(X_{L;q}^{i})+2\Re \mc S_R(X_{R;q}^{i})])\\
    \times&\exp(\kappa\sum_{q=1}^n\sum_i \Tr[X_{L;q}^{i}V_qX_{R;q}^{i}V_q^\dagger+X_{L;q+1}^{i}U_qX_{R;q}^{i}U_q^\dagger]).
\end{split}
\end{equation}
The integrand of $Z_n$, when positive and normalized, may be interpreted as a probability distribution over $n$ copies of the set $\{X^i_L, X^i_R\}$, the correlation functions of which are computed by an `t Hooft expansion. 
The only difference in the additional diagrammatic rule is that each propagator must come with a label $``q"$ to keep track of which copy $X_{L,R;q}^{i}$ the propagator belongs to, as depicted in Fig. \ref{fig:colored-ribbons-bare}. This sequence of $q$ from $1$ to $n$ winds once around the replica circle.

It is important to note that when viewed as a graph (shrinking each ribbon into a line), edges may change label $q \rightarrow q \pm 1$ along their trajectory due to the coupling of $X_{L;q+1}$ to $X_{R;q}$, however interaction vertices, which are contained in $\mc S_{L/R}(X^i_{L/R;q})$, only connect edges of the same label, $q$. In this sense a definite label $q$ can be associated to each vertex. If we take some path $\gamma$ along the graph, then the sequence of vertices along this path will produce an ordered sequence of labels $\sigma_{\gamma} = \{q_1, \ldots, q_{|\gamma|}\}$. We may now classify $\gamma$ based on the winding number of this sequence around the replica circle --- we say that any cycle with nontrivial winding number `wraps the replica circle', and any cycle with trivial winding number `does not wrap the replica circle'. We classify `t Hooft ribbon diagrams in a similar manner --- we say that any ribbon diagram wraps the replica circle if and only if it contains a path $\gamma$ that wraps the replica circle.

We now ask which diagrams contribute to the von Neumann entropy, defined in \eqref{eqn:von-Neumann-replica}.
As mentioned above, we must analytically continue $Z_n$ in $n$ after computing it for all integer $n$. We cannot do this exactly. However, we may diagrammatically expand $Z_n$ into contributions from different families of `t Hooft diagrams. Let $\mathcal{R}^{\vee}_n$ be the set of connected ribbon diagrams that wrap the replica circle with $n$ sites, and let $\mathcal{R}^{\wedge}_n$ be the set of connected ribbon diagrams that do \textit{not} wrap the replica circle. For any diagram $\Gamma$ let $f(\Gamma)$ denote its evaluation as a Feynman diagram. $\log Z_n$ then generates the expansion
\begin{equation}
\log Z_n = \sum_{\Gamma \in \mathcal{R}_n^{\wedge}} f(\Gamma) + \sum_{\Gamma \in \mathcal{R}_n^{\vee}} f(\Gamma). 
\end{equation}
For the formal continuation defined by lifting nonwinding diagrams to the infinite replica chain at fixed local weights, translation symmetry gives
\begin{equation}
\sum_{\Gamma \in \mathcal{R}^{\wedge}_n} f(\Gamma) = n\sum_{\Gamma \in \mathcal{R}_1^{\wedge}}f(\Gamma).
\end{equation}
We can now read off from \eqref{eqn:von-Neumann-replica} that all diagrams in $\mathcal{R}_n^{\wedge}$ cancel in their contribution to the von Neumann entropy. We may therefore write, as a formal asymptotic series,
\begin{equation}
S_{\text{vN}} = -\lim_{n \rightarrow 1} \partial_n \left(\sum_{\Gamma \in \mathcal{R}^{\vee}_n} f(\Gamma) - n\sum_{\Gamma \in \mathcal{R}^{\vee}_1} f(\Gamma) \right).
\end{equation}

We conclude that just as in the case of continuum worldsheets as in \cite{Susskind:1994sm}, only worldsheets with some nontrivial topology in Euclidean time may contribute nontrivially to the entanglement entropy. We do not get to choose which diagrams to include --- we must include \textit{all} diagrams, including those with vortices due to propagators running around the replica circle (just like that in Fig. \ref{fig:colored-ribbons-bare}). In fact, it is precisely such `punctured' diagrams that we may take as the non-perturbative definition of the Susskind-Uglum worldsheets punctured by a conical deficit in the bulk. 

Away from any regime where a smooth geometry is manifest, the behavior of the above punctured 't Hooft diagrams provide a defining signature of the topology of an emergent string target space. The `t Hooft genus counting argument na\"ively suggests punctured diagrams contribute at $O(N^2)$. However, it may happen that gauging $U(N)$ renders diagrams to be suppressed by factors of $N$ as we will soon see. 
We therefore take the following interpretation: if a typical ribbon diagram containing a cycle with nontrivial winding contributes at $O(N^2)$ then the replica circle is contractible in the bulk. Otherwise, we determine that it is \textit{not} contractible, and there is a topological obstruction.

Let us now illustrate the role of gauging in determining contractibility. We choose to gauge-fix the $U_q$ and $V_q$ integrals so that\footnote{At this point $\beta$ is simply an arbitrary constant; we insert it here to uniformize notation and normalization with Appendix \ref{app:BMN-lattice-deconfinement}.} (see e.g. Appendix \ref{app:BMN-lattice-deconfinement} for details) 
\begin{equation}
    V_q=\mathbb{1}~,\qquad\qquad U_q=\Theta=\text{diag}\left(e^{i\beta\theta_a/n}\right)~,\qquad\forall\;q~.
\end{equation}
Propagators and interaction terms now come equipped with insertions of phases, which are then averaged over. For example:
\begin{equation}\label{eqn:gauged-interaction-phases}
\Tr[X^{i}_{L;q+1}\Theta X^{i}_{R;q}\Theta^{\dag}] = \sum_{ab}\left(X^{i}_{L;q+1}\right)_{ab}\left(X^{i}_{R,q}\right)_{ba}e^{i \beta (\theta_b - \theta_a)/n}~.
\end{equation}
The net effect is that any color line carrying color charge $a$ that winds around the replica circle will pick up an overall phase factor of $e^{i \beta\theta_a}$ or $e^{- i\beta\theta_a}$ depending on the direction it winds.

In other words, the fate of an `t Hooft diagram depends on the behavior of expectation values of Polyakov loop insertions of the form
\begin{equation}\label{eqn:general-unitary-traces}
\begin{split}
&\left\langle \prod_{i=1}^{\ms n} \Tr[U^{n_i}] \prod_{j=1}^{\ms m} \Tr[(U^{\dag})^{m_j}] \right\rangle = \sum_{\{a_i, b_j\}}\left\langle \exp(i\beta \sum_{i=1}^{\ms n} n_i \theta_{a_i} - i\beta \sum_{j=1}^{\ms m} m_j \theta_{b_j}  ) \right\rangle~,
\end{split}
\end{equation}
where  we must have $\sum_{i=1}^{\ms n}n_i = \sum_{j=1}^{\ms m} m_j.$

The behavior of the above expectation values is precisely determined by the deconfinement transition \cite{Wadia:1980cp,Gross:1980he,Aharony:2005bq,Alvarez-Gaume:2005dvb,Hadizadeh:2004bf,Kawahara:2007fn,Kawahara:2007ib}. This transition is governed by the effective potential of the unitary degrees of freedom, $U$, upon integrating out the bosonic `matter' fields $X^i$ and controlled by the coupling, $\kappa$. At the sphere topology level we may restrict our analysis to the saddle point eigenvalue distribution of $U$ which we denote $\rho_0(\beta \theta)$:
\begin{equation}\label{eqn:general-unitary-traces-saddle}
\begin{split}
\frac{1}{N^{{\ms n} + {\ms m}}}\left\langle \prod_{i=1}^{\ms n} \Tr[U^{n_i}] \prod_{j=1}^{\ms m} \Tr[(U^{\dag})^{m_j}] \right\rangle \approx &\beta^{\ms m + \ms n} \int_0^{\frac{2\pi}{\beta}} \prod_i  \dd \theta_i\,\rho_0(\beta \theta_i)\prod_j \dd \theta_j\,\rho_0(\beta \theta_j)\\
& \qquad\quad\times\exp(i \beta \sum_i^{\ms n} n_i \theta_i - i \beta \sum_j^{\ms m} m_j \theta_j)\\
&\qquad\qquad\quad+ O(1/N). 
\end{split}
\end{equation}
Below the critical coupling of the deconfinement transition $\kappa_c$, $\rho_0$ is uniform in $\theta$ and the expectation values \eqref{eqn:general-unitary-traces} vanish to leading order in $N$:
\begin{equation}
    \frac{1}{N^{\ms{n} + \ms{m}}}\left\langle\prod_i^{\ms n} \Tr[U^{n_i}] \prod_j^{\ms m} \Tr[(U^{\dag})^{m_j}] \right\rangle \approx O(1/N)~,\qquad \kappa<\kappa_c~.
\end{equation}
Above the deconfinement transition, however, the central $U(1)$ symmetry is spontaneously broken and the eigenvalue distribution becomes nonuniform around some phase $e^{i \beta\theta_0}$ \cite{Wadia:1980cp,Gross:1980he,Aharony:2003sx,Aharony:2005bq}. The exact phase doesn't matter, as $\theta_0$ cancels out of \eqref{eqn:general-unitary-traces-saddle} due to the condition $\sum_{i}n_i = \sum_j m_j$:
\begin{equation}
   \frac{1}{N^{\ms{n}+\ms{m}}} \left\langle\prod_i^{\ms n} \Tr[U^{n_i}] \prod_j^{\ms m} \Tr[(U^{\dag})^{m_j}] \right\rangle \approx O(1)~,\qquad \kappa>\kappa_c~.
\end{equation}
This $O(1)$ estimate requires the winding moments appearing in the correlator to be nonzero.

This is the essential connection between the Polyakov loop order parameter, the proliferation of worldsheet vortices in the `t Hooft diagrams, and the topology of the thermal circle as described by \cite{Atick:1988si} and further developed in \cite{FuruuchiThermal2005,Furuuchi:2006st,FuruuchiLectures2006}; here we explicitly connect it to a transition from $O(1)$ to $O(N^2)$ entanglement entropy. In \S\ref{sec:thermofield-double} we further relate this transition to the existence of Susskind-Uglum edge modes, and to the physics of replica wormholes in \S\ref{sec:replica-wormholes}. 
\begin{figure}[h!]
\centering
    \begin{subfigure}[b]{0.4\textwidth}
        \centering
        \includegraphics[width=\textwidth]{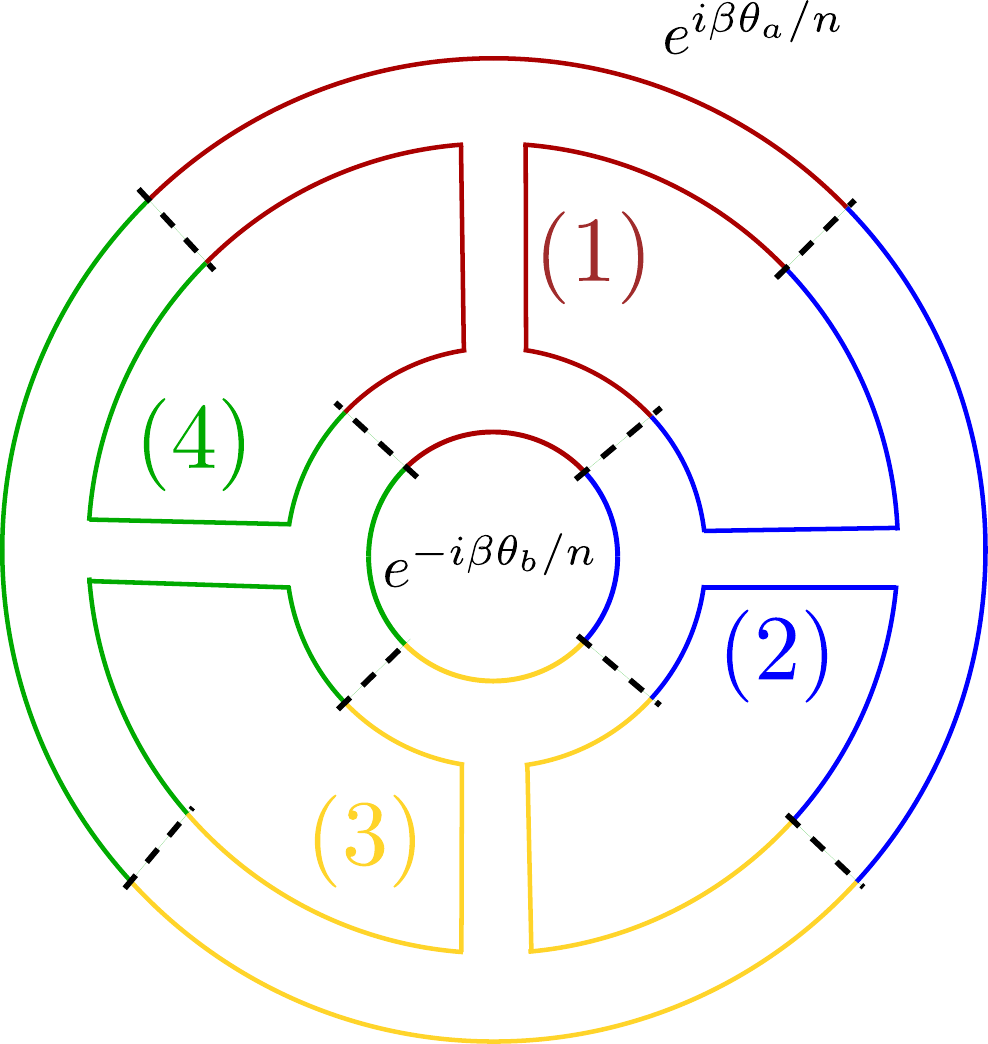}
    \end{subfigure}%
    ~ 
    \begin{subfigure}[b]{0.55\textwidth}
        \centering
        \begin{tikzpicture}[scale=2.5]
            \begin{scope}[rotate=42]
            \draw [dashed, opacity=0.5] (1,0) arc [x radius=1, y radius=0.25, start angle=0, end angle=180];
            \draw [thick, dashed] (1,0) arc [x radius=1, y radius=0.25, start angle=0, end angle=-180];
            \end{scope}
    \begin{scope}[rotate=138]
      \draw [dashed, opacity=0.5] (1,0) arc [x radius=1, y radius=0.5, start angle=0, end angle=180];
      \draw [thick, dashed] (1,0) arc [x radius=1, y radius=0.5, start angle=0, end angle=-180];
    \end{scope}
    \begin{scope}
      \clip [rotate=42] (1,0) arc [x radius=1, y radius=0.25, start angle=0, end angle=-180] -- (-1,-1) -- (1,-1) -- (1,0);
      \clip [rotate=138] (1,0) arc [x radius=1, y radius=0.5, start angle=0, end angle=-180] -- (-1,1) -- (1,1) -- (1,0);
      \shade [ball color = yellow, opacity = 0.7] (0,0) circle [radius=1];
    \end{scope}
    \begin{scope}
      \clip [rotate=42] (1,0) arc [x radius=1, y radius=0.25, start angle=0, end angle=-180] -- (-1,1) -- (1,1) -- (1,0);
      \clip [rotate=138] (1,0) arc [x radius=1, y radius=0.5, start angle=0, end angle=-180] -- (-1,-1) -- (1,-1) -- (1,0);
      \shade [ball color = red, opacity = 0.7] (0,0) circle [radius=1];
    \end{scope}
    \draw[thick,fill=white] (-.5,-.18) ellipse (.06 and .06);
    \node at (-.5,0) {$e^{i\beta\theta_a/n}$};
    \draw[very thick] (-.5,-.18) -- (-1,-.36);
    \draw[very thick, dashed] (-.5,-.18) -- (.5,.18);
    \begin{scope}
      \clip [rotate=138] (1,0) arc [x radius=1, y radius=0.5, start angle=0, end angle=-180] -- (-1,1) -- (1,1) -- (1,0);
      \clip [rotate=42] (1,0) arc [x radius=1, y radius=0.25, start angle=0, end angle=-180] -- (-1,1) -- (1,1) -- (1,0);
      \shade [ball color = green, opacity = 0.5] (0,0) circle [radius=1];
    \end{scope}
    \begin{scope}
      \clip [rotate=138] (1,0) arc [x radius=1, y radius=0.5, start angle=0, end angle=-180] -- (-1,-1) -- (1,-1) -- (1,0);
      \clip [rotate=42] (1,0) arc [x radius=1, y radius=0.25, start angle=0, end angle=-180] -- (-1,-1) -- (1,-1) -- (1,0);
      \shade [ball color = white!20!blue, opacity = 0.7] (0,0) circle [radius=1];
    \end{scope}
    \draw[thick,fill=white] (.5,.18) ellipse (.06 and .06);
    \draw[very thick] (.5,.18) -- (1.25,.45);
    \node at (.5,-.1) {$e^{-i\beta\theta_b/n}$};
\end{tikzpicture}
    \end{subfigure}
    \caption{Two sphere-topology ribbon diagrams that picks up an overall factor of $\sum_{ab}e^{i\beta(\theta_a - \theta_b)}$ stemming from two color lines that wrap the replica circle in opposite directions. \textbf{(Left)} A simple diagram of the same structure as Fig. \ref{fig:colored-ribbons-bare} with relative phases included. \textbf{(Right)} A diagram of the same relative topology as the left, consisting of many vertices and `zoomed out' so that the worldsheet appears as a smooth punctured sphere. The thick black line is the axis of replica symmetry drawn here to make manifest how it `punctures' the sphere.}\label{fig:colored-ribbons}
\end{figure}
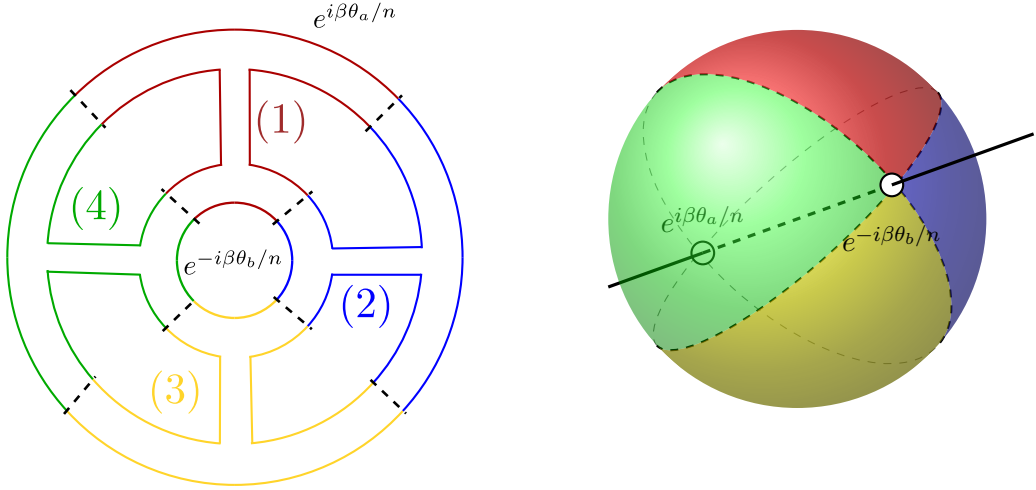

According to the diagnosis in \S\ref{sec:contractible-cycles} for the bulk contractibility of the replica cycle, the deconfinement transition is responsible for the topological transition of the bulk geometry, as probed by closed worldsheets. This physics is illustrated in Fig. \ref{fig:colored-ribbons} --- if the boundary partition function generates punctured sphere diagrams, they can `pass through' the replica circle unobstructed. Otherwise, the replica circle behaves like a topological obstruction. In our point of view the difference in the behavior of the boundary entanglement or thermal entropy is an \textit{implied consequence} of which worldsheet diagrams persist, due to exactly the physical effects depicted in Fig. \ref{fig:Hawking-Page-Worldsheet}.

The above claim is further bolstered by the low-energy interpretation of the resulting $O(N^2)$ contribution to $S_{\text{vN}}$, which matches to a $O(1/G_N)$ `classical' contribution. The Susskind-Uglum computation, and those like it, make clear that such a contribution arises from the Einstein-Hilbert action evaluated over a conical deficit in the bulk. This conical deficit arises precisely if the bulk geometry contains a codimension-two replica fixed locus, i.e. if the $S^1$ cycle winding around the replica circle is contractible in the bulk.

We therefore match three different descriptions of the same physical effect:

\begin{enumerate}

\item In the semiclassical gravitational path integral, the $S^1$ cycle wrapping the boundary replica circle is contractible in the dominant bulk topology.

\item In perturbative string theory, sphere topology worldsheets embedded in an off-shell target space contribute at $O(1/g_s^2)$ due to coming into contact with a conical deficit angle at the replica-symmetric codimension-2 surface in the bulk.

\item In the dual large-$N$ theory, the dominant $O(N^2)$ contribution to the entanglement entropy is computed by `t Hooft diagrams with vortices winding the replica circle.
\end{enumerate}

\section{Susskind-Uglum Open Strings on the Bifurcate Horizon}\label{sec:thermofield-double}

The suggestion of Susskind and Uglum \cite{Susskind:1994sm} is that event horizons (be they black hole or Rindler) partition closed strings into open strings whose endpoints are anchored to the horizon. In another phrasing, the closed string Hilbert space is factorized through an embedding into an extended Hilbert space including open string modes:
\begin{equation}\label{eqn:closed-in-open-x-open}
\mathcal{H}_{\text{closed}} \subset \mathcal{H}_{\text{open}}^\text{in} \otimes \mathcal{H}_{\text{open}}^\text{out}.
\end{equation}
The modern interpretation is that the open string endpoints are edge modes that are responsible for the classical contribution of $A/4 G_N$ to the black hole entropy \cite{Donnelly:2016auv,Donnelly:2020teo}.

The analysis from \S\ref{sec:contractible-cycles} shows how to recover the Euclidean geometry punctured string worldsheets of Susskind and Uglum from MQM wavefunctions and the `t Hooft expansion. We now ask how to recover the corresponding open string degrees of freedom anchored to the bifurcate horizon. Along the way we will need to address the following puzzle: the bosonic degrees of freedom we start with, $X^i$, are adjoints under $U(N)$, whereas we expect that the open string edge modes are labeled by Chan-Paton factors in the fundamental representation. We need a mechanism to find such fundamental degrees of freedom sourcing open `t Hooft worldsheets.

There are two approaches we can take to address this puzzle. In the first approach, described below in \S\ref{sssec:extended-space}, we will extend the Hilbert space and introduce new edge mode degrees of freedom. This is essentially the same as reducing the original $U(N) \cross U(N)$ symmetry to the diagonal $U(N)$. This reintroduces the non-singlet sectors of the two entangled copies of the MQM system under consideration, which by the arguments of \cite{Maldacena:2018vsr} does not spoil the duality to stringy physics. This approach strictly enhances the entanglement, changing the $O(1)$ coefficient in front of $N^2$ term in the entanglement. In the second, alternative, approach described in \S\ref{sssec:no-extended-space}, we do not introduce additional edge mode degrees of freedom. By a series of gauge-fixing and change of variables, we reorganize the purely adjoint degrees of freedom into a system of adjoints and fundamentals belonging to each side. With the original replica sewing and gauge-fixing measure retained, this change of variables does \textit{not} change the entanglement between the two sides.

In what follows we return to the single trace ansatz introduced at the end of \S\ref{ssec:wavefunctions}
\begin{equation}\label{eqn:schematic-wvfnc}
\begin{split}
\Psi_{\kappa
}(X_L^i, X_R^i) &= \sqrt{\mathcal{N}} \exp(-\Tr\mc S_L(X_L^i) - \Tr\mc S_R(X_R^i))\int [\dd U] \exp(\kappa\sum_i\Tr[X_L^i U X_R^i U^{\dag}]),
\end{split}
\end{equation}
where we remain agnostic on the specific form of the potentials $\Tr\mc S_L$ and $\Tr\mc S_R$, apart from them being single trace, or introducing additional intermediate auxiliary variables to integrate out. We simply assume that there is a critical $\kappa_c$ at which the saddle point eigenvalue distribution of the cycle holonomy obtained after replica sewing undergoes a deconfinement transition. The reader is encouraged to keep the simplified BMN thermofield double state introduced earlier, \eqref{eq:TFDbosonicsimplified}, as a specific example in mind.

\subsection{Edge Modes from an Extended Hilbert Space}\label{sssec:extended-space}

First we take the approach of first promoting the $U$ in \eqref{eqn:schematic-wvfnc} to a new degree of freedom by removing the $\int \dd U$ integral from \eqref{eqn:schematic-wvfnc}:
\begin{equation}
\Psi_{\kappa}^\text{ext}(X_L^i, X_R^i, U) = \sqrt{\mathcal{N}_{\rm ext}} \exp(-\Tr\mc S_L(X_L^i) - \Tr\mc S_R(X_R^i) + \kappa\sum_i\Tr[X_L^i U X_R^i U^{\dag}]).
\end{equation}
We then further introduce new degrees of freedom by splitting the bifundamental $U$ into fundamentals belonging to each side by introducing an auxiliary flavor space $\alpha$:
\begin{equation}\label{eqn:U-flavor-decomp}
    U_{ab} = \sum_{\alpha=1}^N\xi_{a \alpha} \eta_{b \alpha}^{*}~,
\end{equation}
and write the wavefunction as
\begin{equation}\label{eqn:Psi-ext-xi-eta}
    \Psi^{\text{ext}}_\kappa(X_L^i, \xi, X_R^i, \eta) = \sqrt{\mathcal{N}_{\rm ext}} \exp(-\Tr\mc S_L(X_L^i) - \Tr\mc S_R(X_R^i) + \kappa\sum_i\sum_{\alpha\beta}\Tr[\xi^{\dag}_\beta X_L^i \xi_\alpha\,\eta^{\dag}_\alpha X_R^i \eta_\beta]).
\end{equation}
Above we view $\xi_{a \alpha}$ as a collection of fundamentals under the left-acting $U(N)$ which conjugates only on $X^i_L$ (similarly for $\eta_{a\alpha}$ and $X^i_R$). The decomposition \eqref{eqn:U-flavor-decomp}, as written, is the unitary factorization $U=\xi\eta^\dagger$ and so viewed as matrices over the $(a,\alpha)$ index pair, $\xi_{a\alpha}$ and $\eta_{a\alpha}$ are also unitary. The frames are integrated with normalized Haar measure, so their columns are constrained to be orthonormal. This decomposition is not unique -- $(\xi,\eta)\mapsto(\xi W,\eta W)$, with $W\in U(N)$, leaves $U$ unchanged; all singular values of $U$ are one. However in principle we could have introduced a larger flavor space with more redundant degrees of freedom. This is essentially the same non-uniqueness one encounters in any extension of a Hilbert space. This extension of the degrees of freedom by splitting $U$ into a product of unitaries is familiar in the study of JT gravity in its formulation as a BF theory and realizing edge modes as quantum mechanics on a group \cite{Lin:2017uzr,Mertens:2022ujr}.

The effect of this decomposition is extending the original Hilbert space to include $L^2$-normalizable functions of $\xi$ and $\eta$:
\begin{equation}
    \ket{\Psi^\text{ext}_\kappa} \in \mathcal{H}_{L} \otimes \mathcal{H}_{\xi} \otimes \mathcal{H}_{\eta} \otimes \mathcal{H}_{R}~.
\end{equation}
Because $\xi$ and $\eta$ interact solely with the left- and right-acting (resp.) $U(N)$, we then treat $\xi$ and $\eta$ as `edge modes' of extended left and right Hilbert spaces:
\beq
    \mc H_{L,\text{ext}}:=\mathcal{H}_{L} \otimes \mathcal{H}_{\xi}~,\qquad\qquad \mc H_{R,\text{ext}}:=\mathcal{H}_{\eta} \otimes \mathcal{H}_{R}~.
\eeq
We may alternatively view this procedure as introducing non-singlet degrees of freedom of $U(N)$ into $\mc H_{L/R}$; this therefore enhances the entanglement. To illustrate this we compute the R\'enyi entropy of $\ket{\Psi^\text{ext}_\kappa}$ between $\mc H_{L,\text{ext}}$ and $\mc H_{R,\text{ext}}$ through replica path integral:
\begin{equation}
\begin{split}
Z_n = \mathcal{N}_{\rm ext}^n \int \prod_{q=1}^n \prod_i[\dd X_{L;q}^{i}][\dd X_{R;q}^{i}] \exp(-2 \Re \sum_{q=1}^n\Tr[\mc S_L(X^{i}_{L;q})+\mc S_R(X^{i}_{R;q})])&\\
\times \int \prod_{q=1}^n [\dd \xi^{(q)}][\dd \eta^{(q)}] \exp(\kappa\sum_{q=1}^n\sum_i\sum_{\alpha\beta}\Tr[\xi^{(q)\dag}_{\beta}X_{L;q}^{i}\xi^{(q)}_{\alpha}\,\eta^{(q) \dag}_{\alpha}X_{R;q}^{i}\eta^{(q)}_{\beta}])&\\
\times \exp(\kappa\sum_{q=1}^n\sum_i\sum_{\alpha\beta}\Tr[\xi^{(q+1)\dag}_{\beta}\,X_{L;q+1}^{i}\xi^{(q+1)}_{\alpha}\eta^{(q) \dag}_{\alpha}X_{R;q}^{i}\eta^{(q)}_{\beta}])&~.
\end{split}
\end{equation}
We now perform a change of variables
\begin{equation}
\left(\tilde{X}^{i}_{L;q}\right)_{\alpha\beta} := \xi^{(q)\dag}_{\alpha}X^i_{L;q}\xi^{(q)}_\beta, \quad \left(\tilde{X}^{i}_{R;q}\right)_{\alpha\beta} := \eta^{(q)\dag}_\alpha X^i_{R;q}\eta^{(q)}_\beta,
\end{equation}
to find
\begin{equation}
\begin{split}
Z_n = \mathcal{N}_{\rm ext}^n \int \prod_{q=1}^n \prod_i[\dd \tilde{X}_{L;q}^{i}] [\dd \tilde{X}_{R;q}^{i}] \exp(-2 \Re \sum_{q=1}^n\tilde{\Tr}[\mc S_L(\tilde{X}^{i}_{L;q})+\mc S_R(\tilde{X}^{i}_{R;q})])&\\
\times \exp(\kappa\sum_{q=1}^n\sum_i\tilde{\Tr}[\tilde{X}_{L;q}^{i}\tilde{X}_{R;q}^{i}] +\kappa \sum_{q=1}^n\sum_i\tilde{\Tr}[\tilde{X}_{L;q+1}^{i}\tilde{X}_{R;q}^{i}])&~. 
\end{split}
\end{equation}
We note that in the above equation, the trace is taken over flavor space and as such we denote it by $\tilde{\Tr}$. This is \textit{exactly} the same $Z_n$ we would compute from the following wavefunction:
\begin{equation}
\tilde{\Psi}_\kappa(\tilde X_L^i, \tilde X_R^i) = \sqrt{\mathcal{N}_{\rm ext}} \exp(- \tilde{\Tr}\mc S_L(\tilde X_L^i) - \tilde{\Tr}\mc S_R(\tilde X_R^i) +\kappa\sum_i\tilde{\Tr}[\tilde X_L^i \tilde X_R^i])~.
\end{equation}
This wavefunction looks precisely like an \emph{un-gauged} version of \eqref{eqn:schematic-wvfnc}, where conjugation has been promoted to a global symmetry, however here we emphasize that this global symmetry is in fact the flavor symmetry. Currently, in writing the decomposition \eqref{eqn:U-flavor-decomp} as a product of unitary frames, the interaction preserves the diagonal flavor $U(N)$ -- we seemed to have found the open string Chan-Paton factors we were looking for. 

We caution again, however, that the decomposition \eqref{eqn:U-flavor-decomp} is not unique: one may choose to introduce redundant degrees of freedom and drastically enhance this flavor symmetry and resulting entanglement entropy. This is simply a reflection of the well-known ambiguity in extending a Hilbert space in the face of gauge-invariance, which is usually cast as an ignorance the true UV degrees of freedom comprising the system. In this case it might be prudent to search for a `minimal' factorization, as advocated for in \cite{Mertens:2025ydx}. However since our current ethos is that large-$N$ matrices provide a non-perturbative definition of the bulk string theory, it would be desirable to eliminate this ambiguity and find the open string edge modes {\it directly} in a given model. This is what we will pursue in the following section.

\subsection{Edge Modes without an Extended Hilbert Space}\label{sssec:no-extended-space}

We now describe how open string edge modes may appear from the original description without extending the Hilbert space by non-singlet degrees of freedom. As we will soon see, the appearance of the edge modes is tied to the deconfinement transition described in \S\ref{sec:contractible-cycles}. To make progress we first gauge-fix the $U(N) \cross U(N)$ symmetry. The idea is that if we diagonalize one of the $X^i_L$ and $X^i_R$ (say $X_{L/R}^{i=1}$ in both cases without loss of generality), the remaining adjoint degrees of freedom $X_{L,R}^i$ with $i \neq 1$ still support a ribbon diagram genus expansion despite the lack of a residual $U(N)$ redundancy. The reason this works is that we do not need all vertices and propagators to contribute with the same value for each color index --- we simply need them all to have the same correct $N$ scaling. This idea has been made precise by Kazakov, Staudacher, and Wynter \cite{Kazakov:1995ae} in the context of background fields. Here we briefly review the idea in schematic form, leaving out the technical details which can be found in \cite{Kazakov:1995ae}.

Consider a $U(N)$ symmetric matrix integral of some adjoint matrices $X^i$. We separate out the quadratic terms explicitly, as these define our propagator:
\begin{equation}\label{eqn:schematic-gauge-fixed-integral}
Z = \int \prod_i [\dd X^i] \exp(-\sum_i \Tr[(X^i)^2] - \Tr[V(X^i)]).
\end{equation}
We then fix the gauge by demanding
\beq
    X^1=\Lambda~,
\eeq
where $\Lambda$ is diagonal. The matrix integral is given by
\begin{equation}
Z = \int [\dd \Lambda]\Delta_\text{V}(\Lambda)\, e^{- \Tr[\Lambda^2]} \int \prod_{i \neq 1}[\dd X^i] \exp(- \sum_{i \neq 1} \Tr[(X^i)^2] - \Tr[V(\Lambda, X^i)])~,
\end{equation}
where $\Delta_\text{V}(\Lambda)=\prod_{a<b}(\lambda_a-\lambda_b)^2$ is the Hermitian eigenvalue Jacobian, with overall gauge-volume and permutation factors absorbed into the measure. We now use the $\int[\dd X^{i\neq 1}]$ integral as the generating function of an `t Hooft expansion, treating $\Lambda$ as a background field; the resulting expansion will then be averaged diagram-by-diagram over the eigenvalue distribution of $\Lambda$. 

The effect of the $\Lambda$ insertions is to generate color-dependent vertices in the polynomial expansion of $\Tr[V(\Lambda, X^i)]$. For example, the common quartic commutator-squared term will produce a term that shifts the propagator: 
\begin{equation}
\Tr[[\Lambda, X^j]^2] = -\sum_{ab} |X^j_{ab}|^2(\lambda_a - \lambda_b)^2.
\end{equation}
More generally we have
\begin{equation}\label{eqn:lambda-dependent-coupling}
\Tr[V(\Lambda, X^i)] = \sum_k \sum_{{a_1 \ldots a_k}}\sum_{i_1, \ldots i_k} X^{i_1}_{a_1, a_2}X^{i_2}_{a_2, a_3}\ldots X^{i_k}_{a_k a_1} f_k(\lambda_{a_1}, \ldots, \lambda_{a_k}),
\end{equation}
where $f_k$ is some polynomial in the $\Lambda$ eigenvalues, $\lambda_a$, into which we have absorbed the coupling constants of the potential. It is these new vertices that we now use to build new ribbon diagrams.\footnote{Note that $k$ can in principle equal 1, creating linear sources for the $X^i$. We define these away by shifting the background to expand around the honest classical $X^i$ saddle for a given distribution $\Lambda$ (although at the lowest genus we may fix $\Lambda$ to the saddle point eigenvalue distribution).}

The genus expansion, at its core, is an order of magnitude argument and so it continues to hold so long as the polynomials $f_k$ scale as 
\begin{equation}\label{eqn:fk-scaling}
f_{k}(\lambda_{a_1}, \ldots \lambda_{a_k}) \sim O(N^{(2 - k)/2})~,
\end{equation}
in the large-$N$ limit. Whether or not this is true depends on the order of magnitude of $\lambda_a$; however indeed for some typical multi-matrix models this order of magnitude is found to be \cite{guionnet2007second}
\begin{equation}\label{eqn:lambda-scaling}
\lambda_a \sim O(\sqrt{N})~,
\end{equation}
identical to the case of a single matrix integral. Because the bare coupling constant for a term like $\Tr[\Lambda^p (X^i)^q]$ has scaling $N^{(2 - p - q)/2}$, when \eqref{eqn:lambda-scaling} is satisfied, this guarantees the necessary scaling of $f_k$, \eqref{eqn:fk-scaling}, needed for a genus expansion.

We caution that while we have presented a generic argument, it may break down in special cases (for example, when the potentials $\Tr[V(X^i)]$ have flat directions that must be lifted by loop effects). For this section, we take \eqref{eqn:lambda-scaling} and \eqref{eqn:fk-scaling} as input assumptions and the subsequent discussions apply to wavefunctions in which they hold --- we have simply reviewed the argument for why these assumptions are not unreasonable. Application of this argument to specific models must be handled on a case-by-case basis.

We will apply this background field method on the gauge invariant wavefunction \eqref{eqn:schematic-wvfnc} (which we rewrite here for convenience):
\begin{equation}
\begin{split}
\Psi_{\kappa
}(X_L^i, X_R^i) &= \sqrt{\mathcal{N}} \exp(-\Tr\mc S_L(X_L^i) - \Tr\mc S_R(X_R^i))\int [\dd U] \exp(\kappa\sum_i\Tr[X_L^i U X_R^i U^{\dag}]),
\end{split}
\end{equation}
assuming \eqref{eqn:lambda-scaling}.
We will also assume that, sufficiently far into the deconfined regime, the $U$ integral localizes around a point $U_0$ maximizing its effectively potential. We can either do so by integrating out $X_{L/R}$, or alternatively, treating $U$ treating $U$ as a function of $X_{L/R}$, maximize $\sum_i\Tr[X^i_L U X^i_R U^{\dag}]$. We expand around this maximizer, $U = U_0 e^{i \epsilon H}$. To order $\epsilon$ the coupling is
\begin{equation}
\begin{split}
&\sum_i\Tr[X^i_L U X^i_R U^{\dag}] = 
\sum_i \Tr[X^i_L U_0 X^i_R U_0^{\dag}] + i \epsilon\sum_i \Tr[U_0^{\dag}X^i_L U_0[H, X^i_R]] + O(\epsilon^2).
\end{split}
\end{equation}
Stationarity requires that $U_0$ (viewed as a function of $X_L^i$ and $X_R^i$) satisfy
\begin{equation}
\sum_i [U_0^{\dag}X_L^i U_0, X_R^i] = 0~.
\end{equation}
Alternatively, we may choose a gauge where $U_0(X_L^i, X_R^i) \equiv \mathbb{1}$ by enforcing the gauge-fixing condition\footnote{This gauge-fixing condition need not pick a unique slice: it also admits other stationary points. We work on a local branch containing the dominant maximum and treat residual stabilizers separately.}
\begin{equation}\label{eqn:commutator-LR-gauge-fixing}
\begin{split}
\sum_i[X_L^i, X_R^i] &= 0~.
\end{split}
\end{equation}
There is a Faddeev-Popov determinant associated to this gauge-fixing and it will be convenient to absorb its square root into the wavefunction so that its inner-product retains the slice constraint, \eqref{eqn:commutator-LR-gauge-fixing}, with no additional measures. We will denote this reweighted wavefunction by $\Psi_\mathrm{gf}$. This can be done by integrating in a set of Grassmannian matrices, $c$ and $\bar c$, and a Hermitian matrix $b$.\footnote{The Grassmannian matrices are familiar in generating Faddeev-Popov determinats, here the auxiliary Gaussian integration over the matrix $b$ combines with the ghost determinant to give the square root \cite{CugliandoloGaussian2004}.}
Working within the gauge, \eqref{eqn:commutator-LR-gauge-fixing}, we parameterize $U = \exp(i H /\sqrt{N})$. Accounting for the Jacobian\footnote{At degree $2m$, $-\log J_\mathrm{Haar}$ is proportional to $N^{-m}\sum_{a<b}(h_a-h_b)^{2m}$. Expanding this sum gives single- and double-trace terms with coefficients $N^{1-m}$ and $N^{-m}$, respectively. These obey the usual $N^{2-r-k/2}$ counting for $r$ traces and total degree $k$, and can contribute at planar order.} on passing from the Haar measure to the flat Hermitian measure
\begin{equation}
    [\dd U]=[\dd H]\,J_\mathrm{Haar}(H)~,\qquad J_\mathrm{Haar}(H)=\prod_{a<b}\left[\frac{\sin\!\left((h_a-h_b)/(2\sqrt N)\right)}{(h_a-h_b)/(2\sqrt N)}\right]^2~,
\end{equation}
where $\{h_a\}$ are the eigenvalues of $H$, we write
\begin{equation}\label{eqn:H-expanded-wavefunction}
\begin{split}
\Psi_\mathrm{gf}(X_L^i, X_R^i) = \sqrt{\mathcal{N}}\exp(-\Tr[\mc S_L(X_L^i)] - \Tr[\mc S_R(X_R^i)] +\kappa \sum_i\Tr[X_L^iX_R^i])&\\
\times \int [\dd H]\,J_\mathrm{Haar}(H) \exp(+\frac{\kappa
}{2N}\sum_i \Tr\Big[[X_L^i,H][X_R^i, H]\Big] + \ldots)&\\
\times \int [\dd b]'[\dd c]'[\dd \bar{c}]' \exp\left(-\frac{\lambda_\text{FP}}{N}\Tr\left[\bar{c}\,\mathcal M(c)+\frac12 b\,\mathcal M(b)\right]\right)&~,
\end{split}
\end{equation}
where $\mathcal M(\cdot)=\sum_i[X_L^i,[X_R^i,\cdot]]$, the primes on $[\dd \cdot]'$ indicate omitting stabilizer zero modes of $\mc M$ (including the central $U(1)$). The arbitrary $\lambda_\text{FP}/N$ coupling is inserted to preserve the `t Hooft limit and only changes a field-independent normalization on the chosen regular branch.

Ellipses contain higher-order couplings in $H$.
Whether or not this expansion is valid precisely depends on the typical ``size'' of $H$, or equivalently the eigenvalue distribution of $U$. We work sufficiently far into the deconfined regime that its integral is localized around $U_0$ (modulo stabilizers). This is an assumption; such localization is realized, e.g., for the thermal holonomy in the high-temperature expansion of \cite{Kawahara:2007fn,Kawahara:2007ib}. The high-temperature expansion of \cite{Kawahara:2007ib} gives a RMS variance of the thermal eigenangle of $\Delta\theta\sim(4N\beta^2/\kappa)^{1/4}$ with $4N\beta^2/\kappa\ll1$ (in our conventions). A pure power of $\kappa$ therefore depends on which parameters are held fixed. For the BMN example, the same scaling additionally requires $\mu(\beta/g^2)^{1/4}\ll1$, so that the quartic commutator potential controls the zero-mode scale.
For the general wavefunctions considered here, we will take as assumption that the matrix configurations dominating the replica integral are localized to $U\sim U_0$, with the typical eigenvalues scaling with a negative power of $\kappa/\beta^2$
\begin{equation}
    \text{spec}(H/\sqrt{N}) = O\left((4N\beta^2/\kappa)^{1/4}\right)~.
\end{equation}
so that \eqref{eqn:H-expanded-wavefunction} furnishes a good approximation to the reweighted wavefunction at some finite truncated order in $H$.

We now parameterize $X_L^i$ and $X_R^i$ by writing
\begin{equation}
    X_{L,R}^1 = V_{L,R} \Lambda_{L,R} V_{L,R}^{\dag}, \qquad X^{i\neq1}_{L,R} = V_{L,R} Y^i_{L,R} V_{L,R}^{\dag}~,
\end{equation}
with $\Lambda_{L/R}$ diagonal and $V_{L/R}$ unitary. The gauge-fixing condition \eqref{eqn:commutator-LR-gauge-fixing} translates to a constraint
\beq\label{eqn:gauge-constraint-UL-explicit}
    [\Lambda_L,(V_L^\dagger V_R)\Lambda_R(V_R^\dagger V_L)]+\sum_{i\neq 1}[Y^i_L,(V_L^\dagger V_R)Y^i_R(V_R^\dagger V_L)]=0~,
\eeq
which is to be imposed on the wavefunction
\begin{equation}\label{eqn:final-gauge-fixed-wavefunction}
\begin{split}
    \Psi_{\mathrm{gf}}&(Y_{L/R}^i, \Lambda_{L/R},V_{L/R})\\&\quad= \exp(-\Tr[\mc S_L(\Lambda_L,Y_L^i)] - \Tr[\mc S_R(\Lambda_R, Y_R^i)]+\kappa \sum_{i \neq 1}\Tr[V_L Y_L^i V_L^{\dag}V_R Y_R^i V_R^{\dag}])\\
    &\quad\quad\quad\times \exp(\kappa \Tr[V_L\Lambda_L V_L^{\dag}V_R \Lambda_R V_R^{\dag}]-\mc V_\text{eff}[\Lambda_{L/R},Y^i_{L/R},V_{L/R}])~.
\end{split}
\end{equation}
Here $\mc V_\text{eff}$ is an effective potential generated by integrating out $b,c,\bar c$, and $H$, with $\mathcal M$ evaluated at $X^{i\neq1}_{L/R}=V_{L/R}Y^i_{L/R}V_{L/R}^{\dag}$ and $X^1_{L/R}=\Lambda_{L/R}$:
\begin{equation}
\begin{split}
\exp\left(-\mc V_\text{eff}\right)=&\int [\dd H]\,J_\mathrm{Haar}(H) \exp(\frac{\kappa}{2N}\sum_i \Tr[[V_LY^i_LV_L^{\dag},H][V_RY^i_RV_R^{\dag},H]] + \ldots)\\
&\quad\times \int [\dd b]'[\dd c]'[\dd \bar{c}]' \exp\left(-\frac{\lambda_\text{FP}}{N}\Tr\left[\bar{c}\,\mathcal M(c)+\frac12 b\,\mathcal M(b)\right]\right)~.
\end{split}
\end{equation}
The most important feature of $\mc V_\text{eff}$ for our purposes is that in the localized regime assumed above, it generates a controlled `t Hooft expansion. The reader is encouraged to compare this wavefunction to \eqref{eqn:Psi-ext-xi-eta}: in the wavefunction \eqref{eqn:final-gauge-fixed-wavefunction}, its expectation values and `t Hooft diagrammatics, $V_L$ exactly plays the role of $\xi$ and $V_R$ plays the role of $\eta$ in \eqref{eqn:Psi-ext-xi-eta}. These are now the fundamentals that source open string `t Hooft diagrams for expectation values computed on either side, which are stitched together with propagators and vertices involving the $b,c$, and $H$ variables.

Wavefunction \eqref{eqn:final-gauge-fixed-wavefunction} is the final result of this section: it is the source for open string diagrams when reduced to a single side.
Consider correlation functions purely on the left side of the geometry built purely from the $(Y_L^i, \Lambda_L)$ degrees of freedom. Integrating out the $(\Lambda_R,Y^i_R,V_R)$ and $V_L$ degrees of freedom will generate non-singlet sources due to the coupling of $V_R$ to $Y_L$ and $\Lambda_L$. These sources generate, in principle, all possible open string topologies with even numbers of boundaries, starting from the annulus, in the gauge-fixed `t Hooft expansion defined by the interaction vertices \eqref{eqn:lambda-dependent-coupling}.

We have therefore found the same type of open-string edge mode structure as the extended Hilbert space picture in \S\ref{sssec:extended-space} purely through a change of variables, with no need to modify the Hilbert space and add additional unphysical entanglement to the system. Notably, this was \textit{only} possible in the deconfined phase of the wavefunction where the series expansion in $H$ as in \eqref{eqn:H-expanded-wavefunction} can be truncated as a trustworthy approximation. This is yet another signature that the `t Hooft diagrammatics knows about the connected wormhole throat --- we do not expect open string edge modes for two copies of entangled near-vacuum AdS, as there is no horizon connecting them. The open string diagrams generated by one-side correlation functions of \eqref{eqn:final-gauge-fixed-wavefunction} in the deconfined regime are then interpreted as being associated to (i.e. living on) the bifurcate horizon of the connected geometry.

In what sense is this compatible with the traditional edge mode picture \eqref{eqn:closed-in-open-x-open}? The essential difference between \eqref{eqn:final-gauge-fixed-wavefunction} and \eqref{eqn:Psi-ext-xi-eta} is that we have coupled the system in \eqref{eqn:Psi-ext-xi-eta} to some auxiliary eigenvalue degrees of freedom, as well as an average over $H,b,c,\bar c$. Let us imagine that the eigenvalue degrees of freedom, $\Lambda_{L/R}$, in \eqref{eqn:final-gauge-fixed-wavefunction} are heavy, in the sense that they are localized to a saddle point eigenvalue distribution with only subleading in $1/N$ (i.e. $g_s$) corrections --- in a putative string dual these would be D-brane fluctuations. The equation \eqref{eqn:closed-in-open-x-open} holds in \eqref{eqn:final-gauge-fixed-wavefunction} as an effective low-energy statement (quite similar to the sense of \cite{Maldacena:2018vsr}) --- at low energies near the bifurcate horizon an observer sees empty space, which may only support closed strings which are then `cut' by the horizon or entangling surface. At high energies brane-like degrees of freedom can fluctuate off of the entangling surface and support open string diagrams.

\section{Replica Wormholes}\label{sec:replica-wormholes}

In this final section we apply the lessons and techniques developed in previous sections for diagnosing bulk topology change in the description of black hole evaporation. To be specific, the exchange of dominance of bulk topology in the gravitational path integral is a crucial aspect in the computation of the unitary Page curve of black hole entropy \cite{Penington:2019kki,Almheiri:2019qdq}. We will briefly review the relevant aspects of the physics before describing how we will model this system.

\begin{figure}[ht]
\centering
    \begin{subfigure}[b]{0.4\textwidth}
        \centering
        \begin{tikzpicture}[scale=.8]
    \def\rad#1#2{\fill[black!40!cyan!40!white,,rounded corners=#1*2mm] (#2) +($(0:#1*2+#1*rnd)$)
        \foreach \a in {20,40,...,350} {  -- +($(\a: #1*2+#1*rnd)$) } -- cycle;}
    \draw[thick,dashed] (0,-3.5) circle (3.5);
    \fill[smooth,magenta!40!white] (1,0) to[out=-110,in=0] (0,-2) to[out=180,in=-70] (-1,0);
    \draw[thick] (1,0) to[out=-110,in=0] (0,-2) to[out=180,in=-70] (-1,0);
    \fill[magenta!40!white] (0,0) ellipse (1 and .5);
    \draw[thick] (0,0) ellipse (1 and .5);
    \begin{scope}[xshift=3.5cm, yshift=-3.5cm,rotate=-90]
        \fill[smooth,magenta!40!white] (1,0) to[out=-110,in=0] (0,-2) to[out=180,in=-70] (-1,0);
        \draw[thick] (1,0) to[out=-110,in=0] (0,-2) to[out=180,in=-70] (-1,0);
        \fill[magenta!40!white] (0,0) ellipse (1 and .5);
        \draw[thick] (0,0) ellipse (1 and .5);
    \end{scope}
    \begin{scope}[xshift=0cm,yshift=-7cm,rotate=180]
        \fill[smooth,magenta!40!white] (1,0) to[out=-110,in=0] (0,-2) to[out=180,in=-70] (-1,0);
        \draw[thick] (1,0) to[out=-110,in=0] (0,-2) to[out=180,in=-70] (-1,0);
        \fill[magenta!40!white] (0,0) ellipse (1 and .5);
        \draw[thick] (0,0) ellipse (1 and .5);
    \end{scope}
    \begin{scope}[xshift=-3.5cm,yshift=-3.5cm,rotate=90]
        \fill[smooth,magenta!40!white] (1,0) to[out=-110,in=0] (0,-2) to[out=180,in=-70] (-1,0);
        \draw[thick] (1,0) to[out=-110,in=0] (0,-2) to[out=180,in=-70] (-1,0);
        \fill[magenta!40!white] (0,0) ellipse (1 and .5);
        \draw[thick] (0,0) ellipse (1 and .5);
    \end{scope}
    \begin{scope}[xshift=-2.5cm,yshift=-1cm]
        \rad{.4}{0,0};
    \end{scope}
    \begin{scope}[xshift=2.5cm,yshift=-1cm]
        \rad{.4}{0,0};
    \end{scope}
    \begin{scope}[xshift=-2.5cm,yshift=-6cm]
        \rad{.4}{0,0};
    \end{scope}
    \begin{scope}[xshift=2.5cm,yshift=-6cm]
        \rad{.4}{0,0};
    \end{scope}
    \begin{scope}[xshift=0cm, yshift=-1.2cm]
        \fill[green, opacity=.35] (0,0) circle[radius=.45cm];
        \draw[black!60!green] (0,0) circle (.45);
        \draw[black!60!green] (0,0) [partial ellipse = 180:360:.45 and .2];
        \draw[black!60!green,dashed] (0,0) [partial ellipse = 0:180:.45 and .2];
    \end{scope}
\end{tikzpicture}
    \end{subfigure}
    \qquad\qquad 
    \begin{subfigure}[b]{0.4\textwidth}
        \centering
        \begin{tikzpicture}[scale=.8]
    \def\rad#1#2{\fill[black!40!cyan!40!white,rounded corners=#1*2mm] (#2) +($(0:#1*2+#1*rnd)$)
        \foreach \a in {20,40,...,350} {  -- +($(\a: #1*2+#1*rnd)$) } -- cycle;}
    \draw[thick,dashed] (0,-3.5) circle (3.5);
    \fill[magenta!40!white] (1,0) to[out=-110,in=135] (1,-2.5) to[out=-45,in=200] (3.5,-2.5) to (3.5,-4.5) to[out=160,in=45] (1,-4.5) to[out=-135,in=110] (1,-7) to (-1,-7) to[out=70,in=-45] (-1,-4.5) to[out=135,in=20] (-3.5,-4.5) to (-3.5,-2.5) to[out=-20,in=-135] (-1,-2.5) to[out=45,in=-70] (-1,0);
    \draw[thick] (1,0) to[out=-110,in=135] (1,-2.5) to[out=-45,in=200] (3.5,-2.5) to (3.5,-4.5) to[out=160,in=45] (1,-4.5) to[out=-135,in=110] (1,-7) to (-1,-7) to[out=70,in=-45] (-1,-4.5) to[out=135,in=20] (-3.5,-4.5) to (-3.5,-2.5) to[out=-20,in=-135] (-1,-2.5) to[out=45,in=-70] (-1,0);
    \fill[magenta!40!white] (0,0) ellipse (1 and .5);
    \draw[thick] (0,0) ellipse (1 and .5);
    \begin{scope}[xshift=3.5cm, yshift=-3.5cm,rotate=-90]
        \fill[magenta!40!white] (0,0) ellipse (1 and .5);
        \draw[thick] (0,0) ellipse (1 and .5);
    \end{scope}
    \begin{scope}[xshift=0cm,yshift=-7cm,rotate=180]
        \fill[magenta!40!white] (0,0) ellipse (1 and .5);
        \draw[thick] (0,0) ellipse (1 and .5);
    \end{scope}
    \begin{scope}[xshift=-3.5cm,yshift=-3.5cm,rotate=90]
        \fill[magenta!40!white] (0,0) ellipse (1 and .5);
        \draw[thick] (0,0) ellipse (1 and .5);
    \end{scope}
    \begin{scope}[xshift=-2.5cm,yshift=-1cm]
        \rad{.4}{0,0};
    \end{scope}
    \begin{scope}[xshift=2.5cm,yshift=-1cm]
        \rad{.4}{0,0};
    \end{scope}
    \begin{scope}[xshift=-2.5cm,yshift=-6cm]
        \rad{.4}{0,0};
    \end{scope}
    \begin{scope}[xshift=2.5cm,yshift=-6cm]
        \rad{.4}{0,0};
    \end{scope}
    \begin{scope}[xshift=0cm, yshift=-3.5cm]
        \fill[green, opacity=.35] (0,0) circle[radius=.45cm];
        \draw[black!60!green] (0,0) circle (.45);
        \draw[black!60!green] (0,0) [partial ellipse = 180:360:.45 and .2];
        \draw[black!60!green,dashed] (0,0) [partial ellipse = 0:180:.45 and .2];
        \draw[very thick] (0,0) circle (.025);
    \end{scope}
\end{tikzpicture}
    \end{subfigure}
    \caption{A computation of the second R\'enyi entropy, via replica trick, between a black hole (depicted as the pink geometry) and its emitted radiation (depicted as a blue blob). \textbf{(Left)} The Hawking saddle, with a sphere worldsheet (depicted in green) in one of the black hole cigars. Just like in thermal AdS, no sphere topology can wrap the replica circle and the entropy is entirely a one-loop effect. \textbf{(Right)} The replica wormhole saddle with a sphere worldsheet. Varying in an analytically continued replica number introduces a bulk conical deficit which can puncture worldsheets in the interior of the geometry.}
    \label{fig:replica-wormholes}
\end{figure}
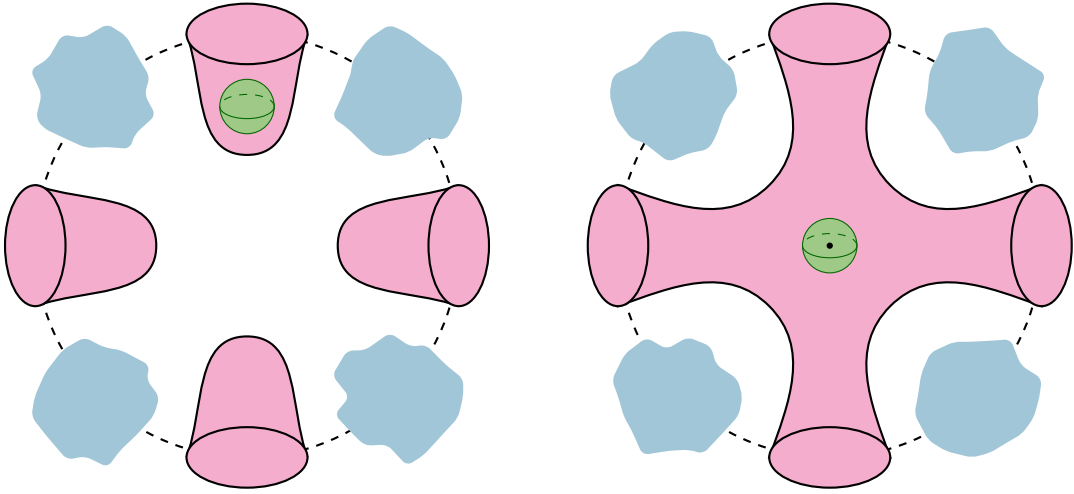

We begin with a spacetime consisting of a gravitating region at small radial coordinate and a non-gravitating region at large radial coordinate. The gravitating region contains a black hole whose radiation we collect. We compute the entanglement entropy of the radiation and the black hole (or, equivalently, of the radiation and the quantum mechanical dual to the black hole) using the replica trick. The crux of the evaluation of the replica trick is that we stick to the gravitational description of the black hole degrees of freedom --- a fluctuating topology and metric. There are two competing saddles --- that in which all the copies of the black hole subsystem remain disconnected, which we call the ``Hawking saddle,'' and that in which they are connected by replica wormholes, the ``replica wormhole saddle.'' These are depicted as cartoon in Fig. \ref{fig:replica-wormholes}. In the entropy evaluation these saddles behave exactly analogously to those involved in the Hawking-Page transition --- the Hawking saddle only contributes a one-loop effect due to field excitations wrapping the replica circle, whereas the replica wormhole saddle produces a conical deficit and its contribution to the entropy is therefore geometric. This is due to the exact same topological effect as we have described above --- the boundary replica circle is not contractible when brought into the bulk in the Hawking saddle, but \textit{is} contractible in the replica wormhole saddle.

Recent discussion in the literature has emphasized that this framework leaves some questions open --- for example, whether a bath prescription necessitates a graviton mass \cite{Geng:2021hlu} (see \cite{Antonini:2025sur} for the counterargument). However, if AdS/CFT duality is a complete theory of the bulk quantum gravity, one must be able to understand the unitary Page curve of a small black hole in AdS with no reference to an auxiliary bath. The evaporation of a small 10-dimensional black hole in AdS$_5 \cross S^5$ must be fully describable by a computation in 4d $ \mathcal{N}=4$ SYM (see \cite{Asplund:2008xd} for a proposed submatrix description of small AdS black holes) with no reference to an auxiliary bath --- both the black hole and the radiation must both be subsystems of the boundary large-$N$ theory.

In the remainder of this section we search for a purely MQM entanglement computation that can recover the exchange of dominance of bulk topology directly from `t Hooft diagrammatics. In a purely quantum mechanical model dual (like MQM) a unitary Page curve of any chaotically evaporating subsystem is manifest --- the puzzle is not in the resolution of the Hawking paradox or the computation of the Page curve, but in the emergence of the bulk geometry and topology dual to the entanglement computation. In other words, if the resolution to Hawking's paradox in non-perturbative string theory is indeed a contribution to the bulk path integral due to strings propagating in a replica wormhole geometry, then we must find a large-$N$ generating functional for the amplitudes of such a string theory. Here we search for such a generating functional in the regime of `t Hooft expansion, using the technology of \S\ref{sec:thermofield-double} to diagnose the transition in the bulk topology.

The strategy in this section is to search for a wavefunction of matrices entangled with some auxiliary system $\Sigma$ whose states we may label by $\sigma$. We will often refer to this system as `the radiation.' This wavefunction is written as $\Psi(X^i,\sigma)$ and lives in the Hilbert space $\mathcal{H}_X\otimes \mathcal{H}_{\Sigma}$. We do not specify the nature of $\Sigma$, but we keep in mind that in the spirit of the preceding discussion we would like the $X^i$ to be some thermal sub-block of BMN or BFSS and $\Sigma$ to be the remaining blocks (as in \cite{Banks:1997hz, Asplund:2008xd, Gautam:2022akq, Berkowitz:2016znt}).

We must choose a schematic ansatz for such a wavefunction. While we are inspired by the arguments in \cite{Berkowitz:2016znt} that search for small black hole configurations (and their dynamical evaporation) in holography, we emphasize that we do \textit{not} prove that the wavefunctions we consider are dual to black holes in some MQM system. Because we do not make reference to a specific Hamiltonian, we do not perform the explicit time evolution of a wavefunction to track an evaporation process and a Page curve. Instead we perform an entanglement computation at fixed time on the boundary and deduce the Euclidean bulk topology probed by the `t Hooft worldsheets as a function of the parameters in our wavefunction ansatz. 

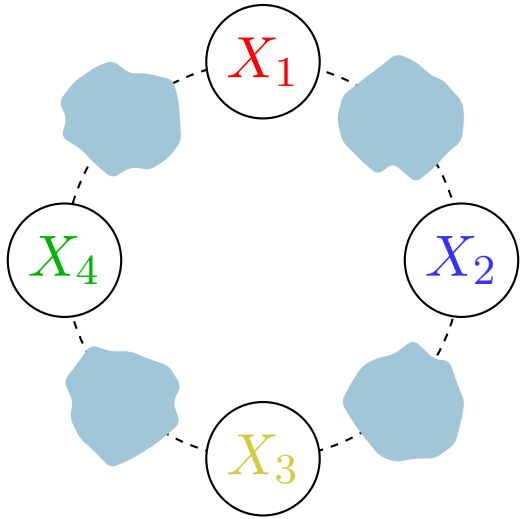
\begin{figure}[ht]
\centering
\begin{tikzpicture}[scale=.75]
        \def\rad#1#2{\fill[black!40!cyan!40!white,rounded corners=#1*2mm] (#2) +($(0:#1*2+#1*rnd)$)
        \foreach \a in {20,40,...,350} {  -- +($(\a: #1*2+#1*rnd)$) } -- cycle;}
    \draw[thick,dashed] (0,-3.5) circle (3.5);
    \fill[white] (0,0) circle (1);
    \draw[thick] (0,0) circle (1);
    \node[red,scale=2] at (0,0) {$X_1$};
    \begin{scope}[xshift=3.5cm, yshift=-3.5cm]
        \fill[white] (0,0) circle (1);
        \draw[thick] (0,0) circle (1);
        \node[white!20!blue,scale=2] at (0,0) {$X_2$};
    \end{scope}
    \begin{scope}[xshift=0cm,yshift=-7cm]
        \fill[white] (0,0) circle (1);
        \draw[thick] (0,0) circle (1);
        \node[black!20!yellow,scale=2] at (0,0) {$X_3$};
    \end{scope}
    \begin{scope}[xshift=-3.5cm,yshift=-3.5cm]
        \fill[white] (0,0) circle (1);
        \draw[thick] (0,0) circle (1);
        \node[black!30!green,scale=2] at (0,0) {$X_4$};
    \end{scope}
    \begin{scope}[xshift=-2.5cm,yshift=-1cm]
        \rad{.4}{0,0};
    \end{scope}
    \begin{scope}[xshift=2.5cm,yshift=-1cm]
        \rad{.4}{0,0};
    \end{scope}
    \begin{scope}[xshift=-2.5cm,yshift=-6cm]
        \rad{.4}{0,0};
    \end{scope}
    \begin{scope}[xshift=2.5cm,yshift=-6cm]
        \rad{.4}{0,0};
    \end{scope}
\end{tikzpicture}
\caption{A diagrammatic depiction of the integral computing the fourth R\'enyi entropy of $\Psi(X^i,\sigma)$, which plays the role of the boundary replica circle. The $\sigma$ degrees of freedom, corresponding to the radiation, are visualized as the blue blob, similarly to Fig. \ref{fig:replica-wormholes}. The replica wormhole computations of \cite{Penington:2019kki,Almheiri:2019qdq} predict that the bulk dual to this matrix integral undergoes the exchange in dominance in topology summarized in Fig. \ref{fig:replica-wormholes}. In \S\ref{ssec:RW-early-times} and \S\ref{ssec:RW-late-times} we show how the `t Hooft diagrammatics capture the corresponding bulk topologies in the early- and late-time ansätze.}\label{fig:RW-replica-circle}
\end{figure}

As an example consider the diagram corresponding to the computation of the fourth R\'enyi entropy that has been drawn in Fig. \ref{fig:RW-replica-circle}. The gravitational path integral analysis predicts that past the Page time, when the $X^i$ are nearly maximally entangled with the exterior, in the sense of Hayden and Preskill \cite{Hayden:2007cs}, the $S^1$ cycle corresponding to the bulk geometry dual to this diagram is contractible. It is this fact about the bulk topology that we wish to probe directly from the MQM side.

\subsection{Early Times --- Non-Contractible Cycle}\label{ssec:RW-early-times}

Since we do not commit to a particular solvable Hamiltonian evolution we must choose some description for the evolution of the wavefunction $\Psi(X^i, \sigma)$ in $\mathcal{H}_X \otimes \mathcal{H}_{\Sigma}$. We begin with an initial product state of the form
\begin{equation}
\Psi_\text{init}(X^i,\sigma) = e^{-\Tr\mc S(X^i)}\Phi_1(\sigma).
\end{equation}
We then model evaporation via the application of simple bilinear operators on the $X^i$ and $\sigma$ degrees of freedom. We introduce some polynomial
\beq
    f(X^i)=\prod_i(X^i)^{l_i}~,
\eeq
whose trace defines a simple operator on $\mathcal{H}_X$ and model a single emitted `quantum' as
\begin{equation}
\begin{split}
&e^{-\Tr\mc S(X^i)}\Phi_1(\sigma) \rightarrow e^{-\Tr\mc S(X^i)}\Phi_1(\sigma) + \epsilon e^{-\Tr\mc S(X^i)}\Tr[f(X^i)] \Phi_2(\sigma)~,
\end{split}
\end{equation}
where $\epsilon$ is some small perturbative parameter and $\Phi_2(\sigma)$ is an orthogonal wavefunction to $\Phi_1(\sigma)$ with respect to the norm on $\mc H_\Sigma$, obtained via the application of some simple operator in $\mathcal{H}_{\Sigma}$. The reduced density matrix $\rho$ from integrating out the radiation is given as
\begin{equation}
\rho(X^i, \tilde{X}^i) = e^{-\Tr\mc S(X^i) - \overline{\Tr\mc S(\tilde{X}^i)}}\left(1 + \epsilon^2 \Tr[f(X^i)]\overline{\Tr[f(\tilde{X}^i)]}  \right).
\end{equation}
We now transform this into a single-trace expression by integrating in a unitary:
\begin{equation}
\rho(X^i, \tilde{X}^i) = e^{-\Tr\mc S(X^i) - \overline{\Tr\mc S(\tilde{X}^i)}}\int [\dd U] \left(1 + N \epsilon^2 \Tr[f(X^i) U f(\tilde{X}^i)^{\dag}U^{\dag}]\right).
\end{equation}
We can now compute $\Tr[\rho^n]$ via the replica trick as before:
\begin{equation}\label{eqn:early-replica-renyi}
\begin{split}
Z_n = &\int \prod_{q = 1}^n [\dd X^{i}_q] \exp(-2 \Re \sum_{q=1}^n \Tr\mc S(X^{i}_q))\\
& \times\int \prod_{q = 1}^n [\dd U_{q}] \left(1 + N \epsilon^2 \Tr[f(X^{i}_q)U_{q}f(X^{i}_{q+1})^{\dag}U_{q}^{\dag}]\right).
\end{split}
\end{equation}
More generally, the $U_{q}$ integrals are replaced by more complicated (perhaps multi-trace) insertions of $X^{i}_q$ and $U_{q}$ traces, such as those in \eqref{eqn:gauged-interaction-phases} in Appendix \ref{app:deconfinement-topology}. These will generate terms in the `t Hooft expansion of the form depicted in Fig. \ref{fig:gauged-colored-ribbon-components} so long as $\Tr\mathcal{S}(X^i)$ is in the appropriate `t Hooft limit.

We now apply the holonomy criterion summarized in \S\ref{sec:contractible-cycles}, using $U := \prod_q U_{q}$ as the unitary whose eigenvalues determine the behavior of sphere diagrams. At the beginning stages of evaporation there is no exponential weight for $U_{q}$, only perturbative insertions. The `t Hooft expansion is computed around the trivial, purely Haar random saddle, and therefore in the confined regime. The total holonomy $U$ only deconfines when enough quanta have been radiated -- i.e. enough perturbative single-trace operators inserted into the $U_{q}$ integrals in $Z_n$ -- to backreact on the saddle-point eigenvalue distribution. Assuming each emitted quantum contributes an $O(1)$ term to the holonomy effective action, an $O(N^2)$ number of emissions can compete with its leading $O(N^2)$ term \cite{DiFrancesco:1993cyw}. This is parametrically consistent with the Page regime if the initial black-hole entropy is $O(N^2)$ and each emitted quantum carries $O(1)$ entropy. Unfortunately, without a more concrete model of dynamical evolution of the radiation we cannot fix the exact transition point. We leave this to future work.

\subsection{Late Times --- Contractible Cycle}\label{ssec:RW-late-times}

We now investigate the entanglement calculation at late times. Again, as we do not have a specific Hamiltonian, we will have to posit a reasonable ansatz to model the system. To make progress we appeal to universal features of chaotic evolution. We furthermore do not attempt to compute a precise time at which topology change occurs (or compare to a Page time) --- we leave more precise computations in this vein to future work. We make a key assumption (inspired by \cite{Hayden:2007cs} in the same spirit as \cite{Liu:2020jsv}):
\paragraph{Assumption:} Over the course of evaporation the matrix subsystem $X^i$ becomes maximally entangled with $\Sigma$ over some microcanonical window centered on some `energy' $E$ with width $2\Delta E$, $\mathcal{H}_{E\pm \Delta E} \subset \mathcal{H}_{X}$, with associated projector $\Pi_{E\pm\Delta E}$. Stated in equations,
let 
\begin{equation}\label{eqn:rhoX}
    \rho_X(X_i, \tilde{X}_i) = \int [\dd\sigma]\, \Psi(X_i, \sigma) \bar{\Psi}(\tilde{X}_i,\sigma)~,
\end{equation}
be the reduced density matrix after integrating out the degrees of freedom in $\Sigma$. Our assumption then is that $\rho$ takes the approximate form
\begin{equation}\label{eqn:projector-assumption}
\rho_X(X_i, \tilde{X}_i) \approx \frac{\Pi_{E\pm\Delta E}(X_i, \tilde{X}_i)}{\tr\Pi_{E\pm\Delta E}} + \delta \rho_X~,
\end{equation}
with $\tr[\delta \rho_X] = 0$ and $\tr[\delta \rho^2_X] \ll \tr[\rho^2_X]$. For the R\'enyi calculations below, we additionally assume that, sufficiently late in evaporation, corrections to $\tr(\rho_X^n)$ are small compared with $D^{1-n}$ at each fixed replica number considered, where $D=\tr\Pi_{E\pm\Delta E}$. We will build the projector $\Pi_{E\pm\Delta E}$ explicitly below.

Without a concrete choice of Hamiltonian we must clarify what we mean by the `energy.' We assume a structure for the Hamiltonian of the form
\begin{equation}
\begin{split}
&H = H_X + H_{\text{int}}(X^i,\sigma) + H_{\Sigma},\\
&H_X = \frac{1}{2}\sum_i \Tr[P_i^2] + \Tr[V(X_i)],\\
& H_{\text{int}} \ll H_{X}, \quad H_{\text{int}} \ll H_{\Sigma}~,
\end{split}
\end{equation}
where $P_i$ is the canonical momentum to $X^i$. In other words, $H_X$ is a single-trace Hamiltonian with a proper `t Hooft limit and $H_{\text{int}}$ may be treated perturbatively. As such, at leading order we take the microcanonical ensemble to be with respect to $H_X$, not necessarily $H$. We additionally assume all three terms in the sum are chaotic. 

It is useful to restate these assumptions directly from a replica trick point of view to clarify their power. We begin with the exact expression for the reduced density matrix $\rho_X$, \eqref{eqn:rhoX},
from which we would derive an exact expression for the $n^\text{th}$ R\'enyi entropy, including both the black hole and the radiation, and integrating out the radiation. Assumption \eqref{eqn:projector-assumption} tells us that to leading order we can do this integral simply by finding an expression for $\Pi_{E\pm\Delta E}(X, \tilde{X})$. We will do so first by constructing the spectral density $\Pi_{\bar E}=\delta(\bar E-H_X)$ and then integrating it over the window:
\beq\label{eq:Ewindow-int}
    \Pi_{E\pm\Delta E}=\int^{E+\Delta E}_{E-\Delta E}\dd\bar E\;\Pi_{\bar E}~.
\eeq
In accordance with the framework of \S\ref{ssec:wavefunctions}, we represent this spectral density using a single-trace expression:
\begin{equation}\label{eqn:single-trace-projector}
\Pi_{\bar E}(X,\tilde X) = \int_{-\infty}^{\infty}\frac{\dd T}{2\pi} [\mathcal{D}A]\prod_i[\mathcal{D}X^i]_{X^i(0) = X^i}^{X^i(T)=\tilde{X}_i} e^{i T\bar E} \exp(i\int_0^T \dd t \Tr[\frac{1}{2}(D_t X_i)^2 - V(X_i)])~,
\end{equation}
where the covariant derivative, $D_tX^i$, is given by \eqref{eq:covder}. The normalization appearing below is $\mathcal{N}=(\tr\Pi_{E\pm\Delta E})^{-1}$. 

We now perform the integral over the energy window. To interpret this resulting expression as having a proper `t Hooft expansion, we \textit{first} generate ribbon diagrams by expanding the $X^i$ path integral. These diagrams will have a $T$-dependent value, which then feeds into the $T$ integral which we perform at the end. In fact we will Wick rotate to Euclidean time, $t\rightarrow -i\tau$ and write $T\rightarrow-i\beta$ (or equivalently replace the Fourier transform with an inverse Laplace transform). We then discretize the Euclidean time interval into $\ms L$ time steps just as we did for the BMN thermofield double (see Appendix \ref{app:BMN-continuum} for details):
\begin{equation}
\begin{split}
\rho_X(X^i, \tilde{X}^i) = \mathcal{N} \int_{0^{+} - i \infty}^{0^{+} + i \infty}\frac{\dd \beta}{\pi i\beta}e^{\beta E} \sinh(\beta \Delta E) \int\prod_{\ms k = 1}^{\ms L-1} \prod_i[\dd X^i_{\ms k}]\;e^{-\sum_{\ms k=0}^{\ms L-1}\Tr \mc S_{\beta}(X^i_{\ms k})}&\\
\times\int \prod_{\ms k=0}^{\ms L-1}[\dd U_{\ms k}]\exp\left(\frac{\ms L}{\beta}\sum_{\ms k=0}^{\ms L-1}\sum_i\Tr[X^i_{\ms k}U_{\ms k}X^i_{\ms k+1}U^\dagger_{\ms k}]\right)&+\delta \rho_X~,
\end{split}
\end{equation}
with a single-trace weight
\begin{equation}
\mc S_\beta(X^i_{\ms k}) = \frac{\ms L}{2\beta}\left(\sum_i(X^i_{\ms k})^2 + \sum_i(X^i_{\ms k+1})^2\right)+\frac{\beta}{\ms L}V(X^i_{\ms k})~,
\end{equation}
and where we take $X^i_0\equiv X^i$ and $X^i_{\ms L}\equiv \tilde X^i$.\footnote{With canonical flat measures, each lattice kernel includes the Gaussian prefactor $(\ms L/(2\pi\beta))^{\ms L D_X/2}$, where $D_X$ counts the independent real matrix components before singlet projection. Each replica carries the corresponding factor with $\beta=\beta_q$. These factors remain inside the contour integrals. The continuum limit $\ms L\to\infty$ is understood.}

The $n^\text{th}$ R\'enyi path integral then has the structure
\begin{equation}\label{eqn:lattice-replica}
\begin{split}
Z_n\approx \mathcal{N}^n \int \prod_{q=1}^n \frac{d \beta_q}{\pi i\beta_q} e^{\beta_qE}\sinh(\beta_q\Delta E) \prod_{\ms k = 1}^{\ms L}\prod_i[\dd X^{i}_{\ms k,q}]e^{-\sum_{q=1}^n\sum_{k=1}^{\ms L}\Tr\mc S_{\beta_q}(X^i_{\ms k,q})}&\\
\times\int\prod_{q=1}^n\prod_{\ms k=1}^{\ms L}[\dd U_{\ms k,q}]\exp\left(\sum_{q=1}^n\frac{\ms L}{\beta_q}\sum_{\ms k=1}^{\ms L}\sum_i\Tr[X_{\ms k,q}^iU_{\ms k,q}X^i_{\ms k+1,q}U^\dagger_{\ms k,q}]\right),& 
\end{split}
\end{equation}
where we make the identifications $X^{i}_{\ms L+1,q}:=X^{i}_{1,q+1}$, $X^{i}_{\ms k,n+1}:=X^{i}_{\ms k,1}$ per replica cyclicity. At fixed $\beta_q$ \eqref{eqn:lattice-replica} is a matrix integral generating function for an `t Hooft expansion generating $\beta_q$-dependent diagrams. In a saddle-point approximation, the energy window $E\pm\Delta E$ determines which $\beta_q$ dominates the $\beta_q$ integral, and therefore the effective overall coupling. As result, the diagrammatics of \eqref{eqn:single-trace-projector} and \eqref{eqn:lattice-replica} depend implicitly on this energy scale. We will assume that the model defined by $H_X$ displays a deconfinement transition as its coupling is tuned. The structure of $Z_n$ in \eqref{eqn:lattice-replica} makes clear that the problem essentially reduces to the gauged replica computation, with the BMN example worked through in Appendix \ref{app:BMN-deconfinement}. The computation of von Neumann entropy is dominated by sphere-topology diagrams that wrap the replica circle. If $E$ is sufficiently large the lattice gauge theory is in the deconfined phase, and this entropy scales as $N^2$. The the interpretation of such diagrams is that they are pierced by conical deficit in the dual bulk geometry. Following the criterion in \S\ref{sec:contractible-cycles}, this is the signature of the replica circle becoming contractible in the bulk.

We therefore arrive again at the picture of sphere-topology `t Hooft diagrams punctured by the replica circle, as in Fig. \ref{fig:colored-ribbons}, generated by the matrix integral in \eqref{eqn:lattice-replica}. These punctured spheres now carry a new interpretation: those living at the center of the replica wormhole geometry as in Fig. \ref{fig:replica-wormholes}. We therefore find that the `t Hooft expansion \textit{informs} us that there must be a topological transition in the dual string theory target space computing the $n$th R\'enyi \eqref{eqn:lattice-replica}, and therefore the entanglement entropy --- this is not a topology that we had to add by hand into the bulk. We therefore propose the large-$E$ regime of \eqref{eqn:lattice-replica} as a schema for a non-perturbative definition of string theory in replica wormhole backgrounds, and it would be interesting future work to construct additional checks.

\section{Discussion}\label{sec:discussion}

In this paper we have demonstrated that at least some features of bulk topology can be directly read off from a systematic `t Hooft expansion. This is a stronger statement than a simple match of a bulk and boundary path integral --- the `t Hooft expansion lives entirely in the quantum mechanical boundary dual, but reproduces features of the target space topology of the bulk worldsheet NLSM. As a proof of concept we have focused on the contractibility of a boundary $S^1$ cycle when brought into the bulk and the resulting implications for the physics. 

The benefits of this framework are two-fold. First, we glean insight into the behavior of the gravitational path integral in string theory. In \S\ref{sec:contractible-cycles}, we have found how the saddle point eigenvalue distribution determines the bulk topology seen by the generated `t Hooft diagrams. This is a peek into the non-perturbative behavior of the string field theory (and NLSM) path integrals --- it is the `t Hooft diagram level realization of the idea that the string field theory path integral is an integral over 2d worldsheet theories, complete with a sum over target space topologies. Semiclassically this becomes a sum over the loop-expansion around all relevant saddle points. We emphasize that this idea is not necessarily in tension with the result from tensionless string theory that the contribution from other topologies is contained in the full string theory path integral around a single saddle \cite{Eberhardt:2021jvj}. We expect that a string field built around one topology, consisting of complicated configurations of non-perturbative excitations, may have an equivalent description as a small fluctuation around some other topology.

The same strategy has allowed us to realize the open string picture of entanglement across the bifurcate horizon in the bulk of the thermofield double, as in \S\ref{sssec:no-extended-space}. It is interesting that we have systematically found an expansion for these open string edge modes only in the deconfined regime, when the assumed angular localization makes the expansion in $H/\sqrt N$ in \eqref{eqn:H-expanded-wavefunction} trustworthy, precisely when we anticipate a connected geometry in the bulk. Again, the properties of the `t Hooft expansion agree with our priors for the behavior of the bulk topology. 

The second benefit of our framework is a new tool to build non-perturbative realizations of string theory in new topologies. \S\ref{sec:replica-wormholes} is a simple example of this idea --- the signature of a conical deficit in the bulk (as a result of bulk contractibility) suggests \eqref{eqn:lattice-replica} as a generating functional for strings in a replica wormhole background (depending, of course, in the specific weights $\mathcal{S}_E$ chosen and whether a deconfinement transition is present). The spirit here is very similar to the philosophy of \cite{Thavanesan:2025ibm}, which searches for a holographic dual to de Sitter by studying the requisite properties of `t Hooft diagrammatics. It would be interesting to determine what signatures of the `t Hooft diagrammatics would be necessary for more general Euclidean wormholes, baby universes, or black hole interiors.

\subsection{Future Work}

There are many facets of this program that would be interesting to make precise in future work.
Firstly, it would be desirable to utilize an explicit dynamical Hamiltonian that takes us through the replica wormhole transition described in \S\ref{sec:replica-wormholes}. Such an example is likely essential for exactly comparing the exchange in topology to the Page time, and capturing subleading topologies that are relevant for the behavior of the entropy close to the Page time.

Another facet is a search for an explicit emergent worldsheet theory in the framework that exhibits topology change using the technology \cite{Gopakumar:2024jfq,Gaiotto:2025hjn}. Because we are only interested in target space topology, the topological string theories that have proved tractable in \cite{Gopakumar:2024jfq} may be sufficient. Some progress in this direction has already been made in \cite{Wong:2026ftn}. It is perhaps possible to explicitly derive the quantitative relationship between the colorings of `t Hooft diagrams and an emergent $S^{1}$-valued target space coordinate beyond the observation in \cite{Klebanov:1991qa} that they must be in the same universality class.

Given the essential relationship between replica wormholes and entanglement islands \cite{Penington:2019kki,Almheiri:2019qdq} we would also like to ask whether the observations in \S\ref{sec:replica-wormholes} can be turned into a statement about entanglement islands and the reconstruction of the black hole interior in the string theory dual of matrix quantum mechanics. Perhaps, progress could be made toward a definition of string theory in the black hole interior.

Lastly, it would be interesting to relate the edge mode structure found in \S\ref{sec:thermofield-double} to a statement about the non-perturbative gauge structure of string theory. The single-trace wavefunction (as in \eqref{eqn:schematic-wvfnc}), and its multi-trace representation found upon integrating out $U$, generate two different diagrammatic expansions that describe the exact same state. One expansion is a connected string theory on a connected geometry, and the other consists of correlated diagrams of closed worldsheets living in two disconnected geometries. This appears to be a realization of a set of gauge redundancies in the emergent string theory that allow for topology change, and perhaps the different wavefunction representations considered in this paper are the correct framework for understanding the nonperturbative nature of this gauge redundancy (following ideas from \cite{Jafferis:2021ywg,Gesteau:2024gzf}).

\section*{Acknowledgements}

We thank Elliott Gesteau, Sean A. Hartnoll, Edward Mazenc, Jacob McNamara, Aron C. Wall,  Wayne W. Weng, Gabriel Wong, and Zhenbin Yang for insightful discussions. We additionally thank Elliott Gesteau, Sean A. Hartnoll, Bob Knighton, Thomas Mertens, and Ronak M Soni for very detailed and insightful comments on a draft of this paper. 
JRF is supported by FNRS MISU grant 40024018 ``Pushing Horizons in Black Hole Physics.'' AF is currently supported by the Simons Center for Geometry and Physics. We acknowledge use of LLMs for checking derivations, and the proofreading and editing of this manuscript.

\appendix

\section{Lattice BMN Thermofield Double and R\'enyi Path Integral}\label{app:BMN-deconfinement}

In this appendix we provide details on the latticized BMN wavefunctionm \eqref{eq:TFDasquiver}, as well as the deconfinement transition of its R\'enyi path integral. For completeness we will also include the fermion components\footnote{We follow the Majorana and gamma-matrix conventions of \cite{KomatsuBMN2024}. 
$T$ acts on spinor indices. We use unnormalized coherent states of eight complex fermionic modes per adjoint generator, with measure $d\mu_F(\bar\eta,\eta)=[d\bar\psi\,d\psi]\exp[-\sum_{a=1}^{8}\Tr(\bar\psi^a\psi^a)]$.} to the wavefunction that have been surpressed in \eqref{eq:TFDasquiver}:

\begin{equation}\label{eq:appTFDasquiver}
\begin{split}
\Psi^\text{BMN}_{\ms L}(X^i_L,\bar\psi_L;X^i_R,\psi_R) = \sqrt{\mc N_{\ms L}}\int &\prod_{\ms k=1}^{\ms L-1} \left(\prod_i [\dd X_{\ms k}^i]\,[\dd\mu_F(\bar\psi_{\ms k},\psi_{\ms k})]\right) \prod_{\ms k=1}^{\ms L} [\dd U_{\ms k}]\prod_{\ms k=1}^{\ms L}\mc K_{F,\ms k}\\
&\times\exp\left(-\frac{2N\ms L}{\beta\,g^2}\sum_{\ms k=1}^{\ms L}\sum_{i=1}^9 \Tr[(U_{\ms k} X^i_{\ms k - 1}U_{\ms k}^{\dag} - X^i_{\ms k})^2]\right)\\
&\qquad\times\exp\left(- \sum_{\ms k=0}^{\ms L}w_{\ms k}\frac{N\beta}{2g^2\ms L}\Tr V_{\bmn}(X^i_{\ms k},0)\right)~,
\end{split}
\end{equation}
where again the weights are $w_0=w_{\ms L}=\tfrac12$ and $w_{\ms k}=1$ for $1\leq\ms k\leq\ms L-1$, and the full potential is given by
\begin{equation}\begin{split}\label{eq:appVBMN}
V_{\bmn}(X^i,\psi^{\alpha}) = &-\frac{1}{2}\sum_{i\neq j}[X^i,X^j]^2+i\frac{2\mu}{3}\sum_{i,j,k=1}^3\epsilon_{ijk}X^iX^jX^k\\
&+\left(\frac{\mu}{3}\right)^2\sum_{i=1}^3(X^i)^2+\left(\frac{\mu}{6}\right)^2\sum_{i=4}^9(X^i)^2\\
&-\sum_i\psi^T\gamma_i[X^i,\psi]+\frac{i\mu}{4}\psi^T\gamma_{123}\psi~.
\end{split}
\end{equation}
The fermionic link kernels are
\begin{equation}\label{eq:BMNfermionkernel}
    \mc K_{F,\ms k}=\langle\bar\psi_{\ms k}|e^{-\epsilon\hat H_F(X_{\ms k},\psi_{\ms k})/2}\,\widehat{\mc U}_{\ms k}\,e^{-\epsilon\hat H_F(X_{\ms k-1},\psi_{\ms k-1})/2}|\psi_{\ms k-1}\rangle~,
\end{equation}
where $\epsilon=\beta/(2\ms L)$, $\widehat{\mc U}_{\ms k}|\psi\rangle=|U_{\ms k}\psi U_{\ms k}^\dagger\rangle$, and
\begin{equation}
    \hat H_F(X)=-\frac{N}{g^2}\sum_i\Tr\big(\hat\psi^T\gamma_i[X^i,\hat\psi]\big)+\frac{i\mu\,N}{4g^2}\Tr\big(\hat\psi^T\gamma_{123}\hat\psi\big)~.
\end{equation}

\subsection{Continuum derivation}\label{app:BMN-continuum}

We now describe how \eqref{eq:appTFDasquiver} can be arrived at from discretizing the thermofield double wavefunction.

The thermofield double is prepared by Euclidean path integral from $\tau\in[0,\beta/2)$. In the BMN model this is expressed as
\begin{align}\label{eq:Psibetacontinuum_app}
    \Psi_\beta(X_L,\bar\psi_L;X_R,\psi_R)=\sqrt{\mc N_\beta}\int &[DX^i]_{X(0)=X_R}^{X(\beta/2)=X_L}[D\psi]_{\psi(0)=\psi_R}^{\bar\psi(\beta/2)=\bar\psi_L}[DA]e^{-\int_0^{\beta/2}\dd\tau\,\mathcal L_\bmn}
\end{align}
with 
\beq
    \mathcal L_\bmn=\frac{N}{g^2}\Tr\left[\sum_{i=1}^9(D_\tau X^i)^2+\psi^T D_\tau\psi+V_\bmn\right]~,
\eeq
where
\beq
    D_\tau X^i=\dot X^i-i[A,X^i]~,\qquad D_\tau \psi^\alpha = \dot \psi^\alpha -i[A,\psi^\alpha]~,\label{eq:covder}
\eeq
and $V_\bmn$ is the potential \eqref{eq:appVBMN}. The fermionic measure includes the coherent-state boundary factors defined by the time slicing in \eqref{eq:BMNfermionkernel}; the right copy is understood as the conjugate Hilbert space. In this setup the $U(N)$ symmetry is gauged in Euclidean time.

We now imagine discretizing this path-integral along the Euclidean time direction into a total of $\ms L$ units of time $\epsilon=\frac{\beta}{2\ms L}$ such that $X^i(\tau)=\{X^i_{\ms k}\}_{\ms k=0,\ldots,\ms L}$ with $X^i_0=X_R$ and $X^i_{\ms L}=X_L$ (with $\psi_0=\psi_R$ and $\bar\psi_{\ms L}=\bar\psi_L$ for the fermions). Denoting the parallel transport operator
\beq
    U_\tau=Pe^{i\int_{\tau-\epsilon}^\tau \dd\tau' A(\tau')}
\eeq
it is easy to verify that for small $\epsilon$
\beq
    X^i(\tau)-U_\tau X^i(\tau-\epsilon) U_\tau^\dagger = \epsilon D_\tau X^i(\tau)+O(\epsilon^2)~.
\eeq
Up to normalization of $\Psi_\beta$ we can replace the integration over Hermitian $A(\tau)$ with the Haar integration over $U_\tau$. Discretizing with the measures $[DX^i]\equiv\prod_{\ms k=1}^{\ms L-1}[\dd X^i_{\ms k}]$, $\prod_{\ms k=1}^{\ms L-1}[\dd\mu_F(\bar\psi_{\ms k},\psi_{\ms k})]$ for the fermions, and $[DU]\equiv\prod_{\ms k=1}^{\ms L}[\dd U_{\ms k}]$, and the link kernels \eqref{eq:BMNfermionkernel}, leads to \eqref{eq:appTFDasquiver}.

\subsection{Deconfinement of the R\'enyi path integral}\label{app:BMN-lattice-deconfinement}

In the following closed-chain calculation, $\beta$ denotes the total circumference and $\ms L$ the number of links. We start with the discretized representation of the R\'enyi path integral 
\beq\label{eqn:U-gauge}
    Z_{\ms L}=\int\prod_{\ms k=1}^{\ms L}[\dd X^i_{\ms k}]\prod_{\ms k=1}^{\ms L}[\dd U_{\ms k}]e^{-\sum_{\ms k=1}^{\ms L}\Tr\mc S(X_{\ms k})}\exp\left(\frac{2N\ms L}{\beta g^2}\sum_{\ms k=1}^{\ms L}\sum_i\Tr\left(X_{\ms k}^iU_{\ms k}X^i_{\ms k-1}U_{\ms k}^\dagger\right)\right)~,
\eeq
where \beq\label{eq:Vk} \mc S(X_{\ms k}):=\frac{N\beta}{\ms Lg^2}V_{\bmn}(X^i_{\ms k},0)+\frac{2N\ms L}{\beta g^2}\sum_i(X^i_{\ms k})^2~. \eeq In showing the deconfinement transition in this model we will follow closely \cite{Aharony:2003sx, Hadizadeh:2004bf}. The first step in this procedure is gauge-fixing the integrals over $U_{\ms k}$. A convenient gauge is the `static' and diagonal gauge where
\beq
    U_{\ms k,ab}\equiv U_{ab}=e^{i\frac{\beta}{\ms L}\theta_a}\delta_{ab}~,\qquad \forall {\ms k}~.
\eeq
This is equivalent to the gauge fixing 
\beq
    \pa_\tau A=0~,\qquad (A)_{ab}=\theta_a\,\delta_{ab}
\eeq
 in the continuum description, \eqref{eq:Psibetacontinuum_app}. We compute the Faddeev-Popov determinant in two steps. The first step is setting all $U_{\ms k} = U$, which results in an associated Fadeev-Popov determinant of
 \begin{equation}
\Delta_1(U) = \prod_{n = 1}^{\ms L - 1} \det(1 - e^{- 2 \pi i n / \ms L} \text{ad}_U) = \prod_{a \neq b}\frac{1 - e^{i \beta (\theta_a - \theta_b)}}{1 - e^{i\beta (\theta_a - \theta_b)/{\ms L}}}.
 \end{equation}
The second step is to diagonalize $U$ itself, resulting in the familiar unitary Vandermonde determinant of
\begin{equation}
\Delta_\text{V}(\theta) = \prod_{a \neq b} (1 - e^{i \beta (\theta_a - \theta_b)/{\ms L}}).
\end{equation}
The total result is (up to a $\theta$-independent normalization)
\beq
    \Delta(\theta) = \Delta_1(\theta)\Delta_\text{V}(\theta)= \prod_{a < b}4\sin^2\left(\frac{\beta(\theta_a - \theta_b)}{2}\right)
\eeq

We will work perturbatively around the classical configuration $X^i_{\ms k}=0$. Fourier transforming
\beq
    X_{\ms k}^i=\frac{1}{\sqrt{\ms L}}\sum_{\ms m=1}^{\ms L}\tilde X_{\ms m}^i\,e^{-i\frac{2\pi}{\ms L}\ms m\ms k},\qquad \tilde X_{-\ms m}^i=(\tilde X_{\ms m}^i)^\dagger
\eeq
with mode indices understood modulo $\ms L$ and integration over independent real components, to one-loop order we have
\beq
\begin{split}
    Z_{\ms L}=&\sqrt{\mc N_{\ms L}}\int\prod_{a=1}^N(\beta \dd\theta_a)\Delta(\theta)\prod_{i=1}^9[\dd\tilde X^i]\\
    &\qquad\times\exp\left(-\frac{N\ms L}{\beta g^2}\sum_{\ms m=1}^{\ms L}\sum_{a,b}\sum_{i=1}^3|(\tilde X^i_{\ms m})_{ab}|^2\left(2-2\cos\left(\frac{2\pi}{\ms L}\ms m+\frac{\beta}{\ms L}(\theta_a-\theta_b)\right)+\frac{\beta^2}{\ms L^2}\frac{\mu^2}{9}\right)\right.\\
    &\qquad\qquad\qquad\left.-\frac{N\ms L}{\beta g^2}\sum_{\ms m=1}^{\ms L}\sum_{a,b}\sum_{i=4}^9|(\tilde X^i_{\ms m})_{ab}|^2\left(2-2\cos\left(\frac{2\pi}{\ms L}\ms m+\frac{\beta}{\ms L}(\theta_a-\theta_b)\right)+\frac{\beta^2}{\ms L^2}\frac{\mu^2}{36}\right)\right)~.
\end{split}
\eeq
The bosonic integrals are simple to perform yielding
\beq
\begin{split}
    Z_{\ms L}=\sqrt{\mc N'_{\ms L}}\int\prod_{a=1}^N(\beta \dd\theta_a)\Delta(\theta)\prod_{\ms m=1}^{\ms L}\prod_{a,b}&\left(2-2\cos\left(\frac{2\pi}{\ms L}\ms m+\frac{\beta}{\ms L}(\theta_a-\theta_b)\right)+\frac{\beta^2}{\ms L^2}\frac{\mu^2}{9}\right)^{-3/2}\\
    &\times\left(2-2\cos\left(\frac{2\pi}{\ms L}\ms m+\frac{\beta}{\ms L}(\theta_a-\theta_b)\right)+\frac{\beta^2}{\ms L^2}\frac{\mu^2}{36}\right)^{-3}
\end{split}
\eeq
The $\theta$-independent factors can contribute to the total entropy but do not affect the eigenvalue saddle. This effective action for $\theta$ is given by
\beq\label{eqn:Seff-form-one}
\begin{split}
    S_\text{eff}(\theta)=-\log\Delta(\theta)+\sum_{\ms m=1}^{\ms L}\sum_{a\neq b}&\left(\frac{3}{2}\log\left(2-2\cos\left(\frac{2\pi}{\ms L}\ms m+\frac{\beta}{\ms L}(\theta_a-\theta_b)\right)+\frac{\beta^2}{\ms L^2}\frac{\mu^2}{9}\right)\right.\\
    &\qquad\left.+3\log\left(2-2\cos\left(\frac{2\pi}{\ms L}\ms m+\frac{\beta}{\ms L}(\theta_a-\theta_b)\right)+\frac{\beta^2}{\ms L^2}\frac{\mu^2}{36}\right)\right)~.
\end{split}
\eeq
We use the identity
\begin{equation}
\prod_{\ms m = 1}^{\ms L}\left(2 - 2 \cos (\frac{2 \pi \ms m}{\ms L}  + \frac{\beta x}{\ms L}) + \frac{\beta^2 y^2}{\ms L^2}\right) = 2 \cosh\left[\ms L\, \text{arccosh}\left(1 + \frac{\beta^2 y^2}{2 {\ms L}^2}\right)\right] - 2 \cos(\beta x)
\end{equation}
to rewrite $S_{\text{eff}}(\theta)$ as
\beq\label{eqn:Seff-form-two}
\begin{split}
    S_\text{eff}(\theta)=-\log\Delta(\theta)+&\sum_{a\neq b}\left(\frac{3}{2}\log\left(2 \cosh\left[\ms L \, \text{arccosh}\left(1 + \frac{\beta^2 \mu^2}{18{\ms L}^2}\right)\right] - 2 \cos(\beta(\theta_a - \theta_b))\right)\right.\\
    &\quad\left.+3\log\left(2 \cosh\left[\ms L \, \text{arccosh}\left(1 + \frac{\beta^2 \mu^2}{72{\ms L}^2}\right)\right] - 2 \cos(\beta(\theta_a - \theta_b))\right)\right)~.
\end{split}
\eeq
We now expand to quadratic order in the variables
\begin{equation}
\phi_n := \sum_a e^{i n \beta \theta_a}~.
\end{equation}
Using the expansions
\begin{equation}
\begin{split}
&\sum_{a < b} \log \left[-\frac{1}{4 \pi^2}\left(e^{i\beta(\theta_a - \theta_b)/2} - e^{-i\beta(\theta_a - \theta_b)/2}\right)^2 \right] = \text{const.} - \sum_{n = 1}^{\infty}\frac{|\phi_n|^2}{n},\\
&\sum_{a\neq b} \log (\cosh \eta - \cos \beta (\theta_a - \theta_b)) = \text{const.} - 2\sum_{n=1}^{\infty}\frac{e^{n \eta}}{n}|\phi_n|^2 \quad (\eta < 0)~,
\end{split}
\end{equation}
where the constants are independent of $\phi_n$, we find
\begin{equation}\label{eqn:Seff-final}
S_{\text{eff}}(\theta) = \text{const.} + \sum_{n = 1}^{\infty}\frac{1}{n}\left[1 - 3 (\mathcal{C}_{1,\ms L})^n - 6 (\mathcal{C}_{2, \ms L})^n\right]|\phi_n|^2~,
\end{equation}
where
\begin{equation}
\begin{split}
& \mathcal{C}_{1, \ms L} := \exp(- \ms L \, \text{arccosh}\left(1 + \frac{\beta^2 \mu^2}{18 {\ms L}^2}\right))\\
& \mathcal{C}_{2, \ms L} := \exp(- \ms L \, \text{arccosh}\left(1 + \frac{\beta^2 \mu^2}{72 {\ms L}^2}\right))~.
\end{split}
\end{equation}
When the coefficient of $\abs{\phi_n}^2$ in the effective action is positive, $S_\text{eff}$ has its extremum at $\phi_n=0$; this is the confined phase. However if one of the coefficients obtains a negative value then the uniform eigenvalue distribution becomes unstable. This first occurs for $\phi_1$.

As we tune $\beta$ the coefficients $\mathcal{C}$ are tuned from $0$ to $1$, signaling a deconfinement transition at the point
\begin{equation}\label{eq:app-deconf-criterion}
3\mathcal{C}_{1,\ms L} + 6\mathcal{C}_{2,\ms L} = 1.    
\end{equation} 
We note that this transition exists regardless of the value of $\ms L$ (it only sets point at which \eqref{eq:app-deconf-criterion} is satisfied): the deconfinement transition exists even for $\ms L=1$. We have therefore verified at one loop that the example wavefunction \eqref{eq:TFDasquiver} produces matrix integral expressions for $n$th R\'enyi path integrals that undergo a deconfinement transition, and therefore pass from a phase where the punctured spheres depicted in Fig. \ref{fig:colored-ribbons} vanish to a phase where they carry $O(1)$ prefactors.

\section{General Gauged and Entangled Single-Trace Wavefunctions}\label{app:deconfinement-topology}

In this appendix we give a more general treatment of the gauged replica calculation, the subsequent color changing diagrams, and their holonomies, supplementing \S\ref{sec:contractible-cycles}. 

\begin{figure}[ht]
\centering
\includegraphics[width=0.8\textwidth]{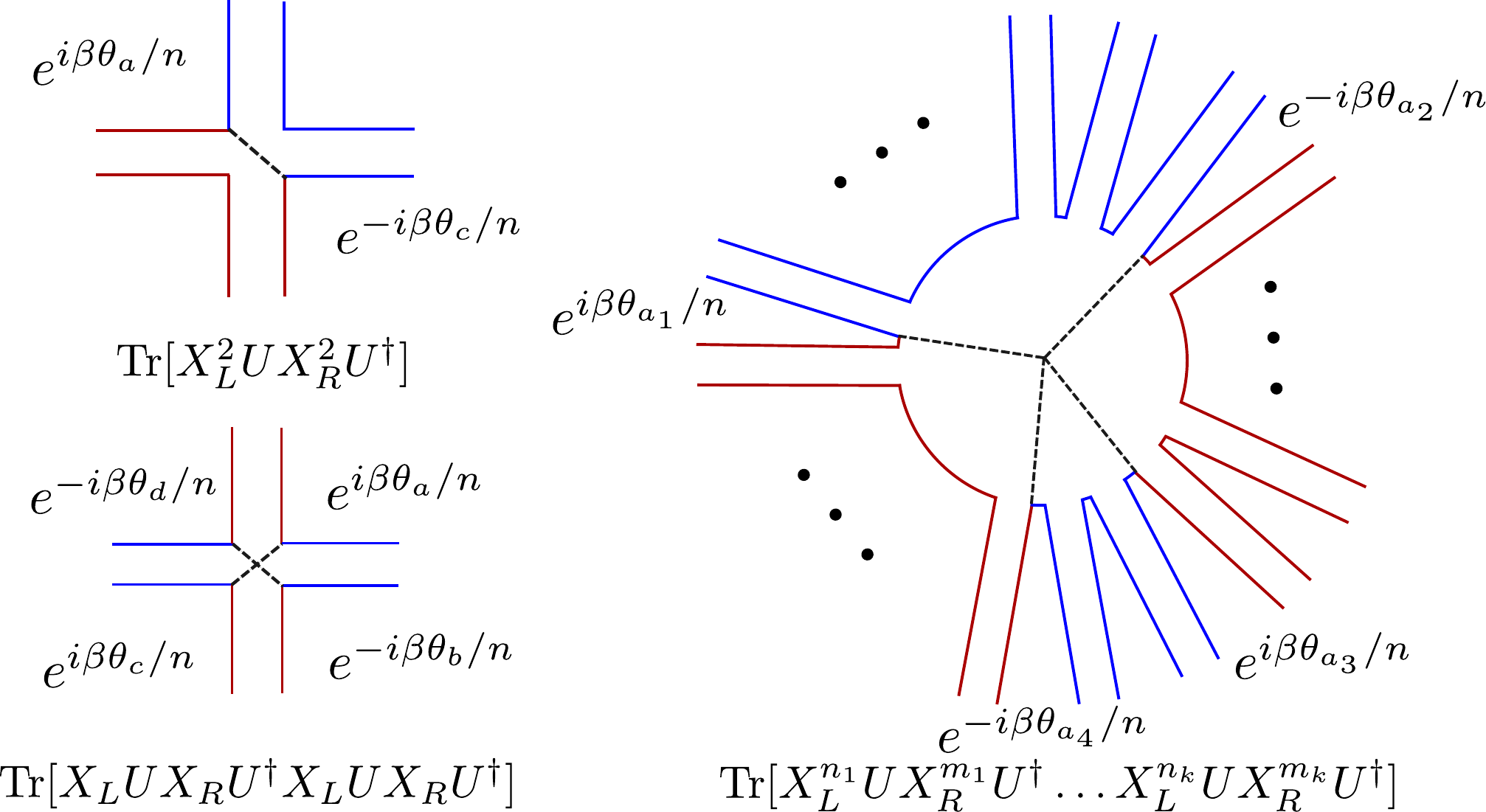}
\caption{Some examples of color-changing interaction vertices generated by interaction terms as \eqref{eqn:general-S-int}. We have already gauge-fixed and diagonalized the unitary $U$, resulting in `t Hooft diagrams weighted by the phases given in \eqref{eqn:gauged-interaction-phases}. A color line (i.e. the edge of a ribbon) that carries color $a$ fully wraps around the replica circle will pick up a total phase of $e^{\pm i \beta \theta_a}$, with the sign depending on the direction of winding (see Fig. \ref{fig:colored-ribbons} for more explicit details.)}\label{fig:gauged-colored-ribbon-components}
\end{figure}

We write a general gauged single-trace allowing a generic interaction term $\mathcal{S}_{\text{int}}$ responsible for generating the entanglement between $X_L^i$ and $X_R^i$: 
\begin{equation}\label{eqn:section-3-wavefunction-gauged}
    \Psi(X_L^i, X_R^i) = \sqrt{\mathcal{N}}\exp(-\Tr[\mc S_L(X_L^i)+\mc S_R(X_R^i)])\int[\dd U]\exp(-\Tr[\mc S_\text{int}(X_L,UX_RU^\dagger)]).
\end{equation}
A general structure for terms that can appear in $\mathcal{S}_{\text{int}}$ can be parameterized as follows. Let the indices $\aleph_s$ run over all possible words (i.e. monomials) that can be made from the symbols $X^i$, and let such words be denoted by $w_{\aleph_s}(X^i)$. $\mathcal{S}_{\text{int}}$ can then be expanded as
\begin{equation}\label{eqn:general-S-int}
\Tr\mathcal{S}_{\text{int}}(X_L, UX_RU^\dagger) = \sum_{\{\aleph_{s}\}} c_{\{\aleph_s\}} \Tr[w_{\aleph_1}(X^i_L) U w_{\aleph_2}(X^i_{R})U^{\dag}w_{\aleph_3}(X^i_L)U\ldots w_{\aleph_s}(X_R^i)U^{\dag}\ldots].
\end{equation} The structure of the integral computing the $n^\text{th}$ R\'enyi entropy is then 
\begin{equation}\label{eqn:nth-Renyi-integral-gauged}
\begin{split}
Z_n = \mathcal{N}^n \int \prod_{q=1}^n\left(\prod_i[\dd X_{L;q}^{i}][\dd X_{R;q}^{i}]\right)[\dd U_q][\dd V_q]   \exp(- \sum_k \Tr[2\Re \mc S_L(X_{L;k}^{i})+2\Re \mc S_R(X_{R;k}^{i})])&\\
\times\exp(-\sum_{q=1}^n \Tr[\mathcal{S}_{\text{int}}(X_{L;q}^i, V_q X_{R;q}^i V_q^\dagger)] - \sum_{q=1}^n\Tr[\overline{\mathcal{S}}_{\text{int}}(X_{L;q+1}^i, U_q X_{R;q}^i U_q^\dagger)])&~.
\end{split}
\end{equation}
As before we gauge-fix the $U_q$ and $V_q$ integrals
\begin{equation}
    V_q=\mathbb{1}~,\qquad\qquad U_q=\Theta=\text{diag}\left(e^{i\beta\theta_a/n}\right)~,\qquad\forall\;q~.
\end{equation}
which leads to phases in propagators and interactions, such as
\begin{equation}\label{eqn:gauged-interaction-phases}
\begin{split}
&\Tr[X^{i}_{L;q+1}\Theta X^{i}_{R;q}\Theta^{\dag}] = \sum_{ab}\left(X^{i}_{L;q+1}\right)_{ab}\left(X^{i}_{R,q}\right)_{ba}e^{i \beta (\theta_b - \theta_a)/n}.\\
&\Tr[X^i_{L;q+1}\Theta X^{i}_{R;q}\Theta^{\dag}X^j_{L;q+1}\Theta X^{j}_{R;q}\Theta^{\dag}]\\
&\qquad\qquad=\sum_{abcd}\left(X^{i}_{L;q+1}\right)_{ab}\left(X^{i}_{R,q}\right)_{bc}\left(X^{j}_{L;q+1}\right)_{cd}\left(X^{j}_{R,q}\right)_{da}e^{i \beta (\theta_b - \theta_c + \theta_d - \theta_a)/n}\\
&\Tr[X^i_{L;q+1} X^j_{L;q+1} \Theta X^i_{R;q} X^j_{R;q}\Theta^{\dag}]\\
&\qquad\qquad=\sum_{abcd}\left(X^{i}_{L;q+1}\right)_{ab}\left(X^{j}_{L;q+1}\right)_{bc}\left(X^{i}_{R,q}\right)_{cd}\left(X^{j}_{R,q}\right)_{da}e^{i \beta (\theta_c - \theta_a)/n}
\end{split}
\end{equation}
Each color-changing vertex is then weighted by an appropriate phase factor when we evaluate the ribbon diagram, as illustrated in Fig. \ref{fig:gauged-colored-ribbon-components}.

\bibliographystyle{JHEP}
\bibliography{refs}

@article{DonnellyWongEBranes2017,
  author = {Donnelly, William and Wong, Gabriel},
  title = {{Entanglement branes in a two-dimensional string theory}},
  journal = {JHEP},
  volume = {09},
  pages = {097},
  year = {2017},
  doi = {10.1007/JHEP09(2017)097},
  eprint = {1610.01719},
  archivePrefix = {arXiv},
  primaryClass = {hep-th}
}

@article{FuruuchiThermal2005,
  author = {Furuuchi, Kazuyuki},
  title = {{From Free Fields to AdS -- Thermal Case}},
  journal = {Phys. Rev. D},
  volume = {72},
  pages = {066009},
  year = {2005},
  doi = {10.1103/PhysRevD.72.066009},
  eprint = {hep-th/0505148},
  archivePrefix = {arXiv}
}

@article{FuruuchiLectures2006,
  author = {Furuuchi, Kazuyuki},
  title = {{Lectures On AdS-CFT At Weak 't Hooft Coupling At Finite Temperature}},
  year = {2006},
  eprint = {hep-th/0608181},
  archivePrefix = {arXiv}
}

@article{HalderJafferisWorldsheet2023,
  author = {Halder, Indranil and Jafferis, Daniel L.},
  title = {{Thermal Bekenstein-Hawking entropy from the worldsheet}},
  year = {2023},
  eprint = {2310.02313},
  archivePrefix = {arXiv},
  primaryClass = {hep-th}
}

@article{Atick:1988si,
    author = "Atick, Joseph J. and Witten, Edward",
    title = "{The Hagedorn Transition and the Number of Degrees of Freedom of String Theory}",
    reportNumber = "IASSNS-HEP-88-14",
    doi = "10.1016/0550-3213(88)90151-4",
    journal = "Nucl. Phys. B",
    volume = "310",
    pages = "291--334",
    year = "1988"
}

@article{Wong:2026ftn,
    author = "Wong, Gabriel",
    title = "{Entanglement and geometric transitions in topological string theory}",
    eprint = "2607.03526",
    archivePrefix = "arXiv",
    primaryClass = "hep-th",
    month = "7",
    year = "2026"
}

@article{Mertens:2025ydx,
    author = "Mertens, Thomas G. and Wu, Qi-Feng",
    title = "{Minimal factorization of Chern-Simons theory {\textendash} Gravitational anyonic edge modes}",
    eprint = "2505.00501",
    archivePrefix = "arXiv",
    primaryClass = "hep-th",
    doi = "10.21468/SciPostPhys.20.4.095",
    journal = "SciPost Phys.",
    volume = "20",
    number = "4",
    pages = "095",
    year = "2026"
}

@article{Gopakumar:2005fx,
    author = "Gopakumar, Rajesh",
    title = "{From free fields to AdS: III}",
    eprint = "hep-th/0504229",
    archivePrefix = "arXiv",
    doi = "10.1103/PhysRevD.72.066008",
    journal = "Phys. Rev. D",
    volume = "72",
    pages = "066008",
    year = "2005"
}

@article{Gopakumar:2004qb,
    author = "Gopakumar, Rajesh",
    title = "{From free fields to AdS. 2.}",
    eprint = "hep-th/0402063",
    archivePrefix = "arXiv",
    doi = "10.1103/PhysRevD.70.025010",
    journal = "Phys. Rev. D",
    volume = "70",
    pages = "025010",
    year = "2004"
}

@article{Gopakumar:2003ns,
    author = "Gopakumar, Rajesh",
    title = "{From free fields to AdS}",
    eprint = "hep-th/0308184",
    archivePrefix = "arXiv",
    doi = "10.1103/PhysRevD.70.025009",
    journal = "Phys. Rev. D",
    volume = "70",
    pages = "025009",
    year = "2004"
}

@article{Bun:2014dha,
    author = "Bun, Joel and Bouchaud, Jean-Philippe and Majumdar, Satya N. and Potters, Marc",
    title = "{Instanton Approach to Large $N$ Harish-Chandra-Itzykson-Zuber Integrals}",
    eprint = "1403.7763",
    archivePrefix = "arXiv",
    primaryClass = "cond-mat.stat-mech",
    doi = "10.1103/PhysRevLett.113.070201",
    journal = "Phys. Rev. Lett.",
    volume = "113",
    pages = "070201",
    year = "2014"
}

@article{Matytsin:1993iq,
    author = "Matytsin, A.",
    title = "{On the large N limit of the Itzykson-Zuber integral}",
    eprint = "hep-th/9306077",
    archivePrefix = "arXiv",
    reportNumber = "PUPT-1405",
    doi = "10.1016/0550-3213(94)90471-5",
    journal = "Nucl. Phys. B",
    volume = "411",
    pages = "805--820",
    year = "1994"
}

@article{Itzykson:1979fi,
    author = "Itzykson, C. and Zuber, J. B.",
    title = "{The Planar Approximation. 2.}",
    reportNumber = "SACLAY-DPh-T 79/82",
    doi = "10.1063/1.524438",
    journal = "J. Math. Phys.",
    volume = "21",
    pages = "411",
    year = "1980"
}

@article{harish1957differential,
  title={Differential operators on a semisimple Lie algebra},
  author={Harish-Chandra},
  journal={American Journal of Mathematics},
  pages={87--120},
  year={1957},
  publisher={JSTOR}
}

@article{Hadizadeh:2004bf,
    author = "Hadizadeh, Shirin and Ramadanovic, Bojan and Semenoff, Gordon W. and Young, Donovan",
    title = "{Free energy and phase transition of the matrix model on a plane-wave}",
    eprint = "hep-th/0409318",
    archivePrefix = "arXiv",
    doi = "10.1103/PhysRevD.71.065016",
    journal = "Phys. Rev. D",
    volume = "71",
    pages = "065016",
    year = "2005"
}

@article{Aharony:2003sx,
    author = "Aharony, Ofer and Marsano, Joseph and Minwalla, Shiraz and Papadodimas, Kyriakos and Van Raamsdonk, Mark",
    editor = "Doebner, H. D. and Dobrev, V. K.",
    title = "{The Hagedorn - deconfinement phase transition in weakly coupled large N gauge theories}",
    eprint = "hep-th/0310285",
    archivePrefix = "arXiv",
    reportNumber = "WIS-29-03-DPP",
    doi = "10.4310/ATMP.2004.v8.n4.a1",
    journal = "Adv. Theor. Math. Phys.",
    volume = "8",
    pages = "603--696",
    year = "2004"
}

@article{Fliss:2025kzi,
    author = "Fliss, Jackson R. and Frenkel, Alexander and Hartnoll, Sean A. and Soni, Ronak M.",
    title = "{Minimal areas from entangled matrices}",
    eprint = "2408.05274",
    archivePrefix = "arXiv",
    primaryClass = "hep-th",
    doi = "10.21468/SciPostPhys.18.6.171",
    journal = "SciPost Phys.",
    volume = "18",
    number = "6",
    pages = "171",
    year = "2025"
}

@article{tHooft:1973alw,
    author = "'t Hooft, Gerard",
    editor = "Taylor, J. C.",
    title = "{A Planar Diagram Theory for Strong Interactions}",
    reportNumber = "CERN-TH-1786",
    doi = "10.1016/0550-3213(74)90154-0",
    journal = "Nucl. Phys. B",
    volume = "72",
    pages = "461",
    year = "1974"
}

@article{Gopakumar:2024jfq,
    author = "Gopakumar, Rajesh and Kaushik, Rishabh and Komatsu, Shota and Mazenc, Edward A. and Sarkar, Debmalya",
    title = "{Strings from Feynman Diagrams}",
    eprint = "2412.13397",
    archivePrefix = "arXiv",
    primaryClass = "hep-th",
    month = "12",
    year = "2024"
}

@article{Gopakumar:2022djw,
    author = "Gopakumar, Rajesh and Mazenc, Edward A.",
    title = "{Deriving the Simplest Gauge-String Duality -- I: Open-Closed-Open Triality}",
    eprint = "2212.05999",
    archivePrefix = "arXiv",
    primaryClass = "hep-th",
    month = "12",
    year = "2022"
}

@article{Susskind:1994sm,
    author = "Susskind, Leonard and Uglum, John",
    title = "{Black hole entropy in canonical quantum gravity and superstring theory}",
    eprint = "hep-th/9401070",
    archivePrefix = "arXiv",
    reportNumber = "SU-ITP-94-1",
    doi = "10.1103/PhysRevD.50.2700",
    journal = "Phys. Rev. D",
    volume = "50",
    pages = "2700--2711",
    year = "1994"
}

@article{Gaiotto:2024dwr,
    author = "Gaiotto, Davide and L{\'o}pez-Raven, Adri{\'a}n and Silverans, Hanne and Zeng, Keyou",
    title = "{Categorical {\textquoteright}t Hooft expansion and chiral algebras}",
    eprint = "2411.00760",
    archivePrefix = "arXiv",
    primaryClass = "hep-th",
    doi = "10.1007/JHEP12(2025)141",
    journal = "JHEP",
    volume = "12",
    pages = "141",
    year = "2025"
}

@inproceedings{Gaiotto:2025hjn,
    author = "Gaiotto, Davide",
    title = "{The Categorical 't Hooft Expansion}",
    booktitle = "{International Congress of Mathematicians 2026}",
    note = "{Submitted to the proceedings}",
    eprint = "2511.19776",
    archivePrefix = "arXiv",
    primaryClass = "hep-th",
    month = "11",
    year = "2025"
}

@article{Brezin:1977sv,
    author = "Brezin, E. and Itzykson, C. and Parisi, G. and Zuber, J. B.",
    title = "{Planar Diagrams}",
    reportNumber = "SACLAY-DPH-T-77-126",
    doi = "10.1007/BF01614153",
    journal = "Commun. Math. Phys.",
    volume = "59",
    pages = "35",
    year = "1978"
}

@article{Witten:1979kh,
    author = "Witten, Edward",
    title = "{Baryons in the 1/n Expansion}",
    reportNumber = "HUTP-79-A007",
    doi = "10.1016/0550-3213(79)90232-3",
    journal = "Nucl. Phys. B",
    volume = "160",
    pages = "57--115",
    year = "1979"
}

@article{Saad:2019lba,
    author = "Saad, Phil and Shenker, Stephen H. and Stanford, Douglas",
    title = "{JT gravity as a matrix integral}",
    eprint = "1903.11115",
    archivePrefix = "arXiv",
    primaryClass = "hep-th",
    month = "3",
    year = "2019"
}

@inproceedings{Klebanov:1991qa,
    author = "Klebanov, Igor R.",
    title = "{String theory in two-dimensions}",
    booktitle = "{Spring School on String Theory and Quantum Gravity (to be followed by Workshop)}",
    eprint = "hep-th/9108019",
    archivePrefix = "arXiv",
    reportNumber = "PUPT-1271",
    month = "7",
    year = "1991"
}

@inproceedings{Ginsparg:1993is,
    author = "Ginsparg, Paul H. and Moore, Gregory W.",
    title = "{Lectures on 2-D gravity and 2-D string theory}",
    booktitle = "{Theoretical Advanced Study Institute (TASI 92): From Black Holes and Strings to Particles}",
    eprint = "hep-th/9304011",
    archivePrefix = "arXiv",
    reportNumber = "YCTP-P23-92, LA-UR-92-3479",
    pages = "277--469",
    month = "10",
    year = "1993"
}

@article{Berenstein:2002jq,
    author = "Berenstein, David Eliecer and Maldacena, Juan Martin and Nastase, Horatiu Stefan",
    title = "{Strings in flat space and pp waves from N=4 superYang-Mills}",
    eprint = "hep-th/0202021",
    archivePrefix = "arXiv",
    doi = "10.1088/1126-6708/2002/04/013",
    journal = "JHEP",
    volume = "04",
    pages = "013",
    year = "2002"
}

@article{Lewkowycz:2013nqa,
    author = "Lewkowycz, Aitor and Maldacena, Juan",
    title = "{Generalized gravitational entropy}",
    eprint = "1304.4926",
    archivePrefix = "arXiv",
    primaryClass = "hep-th",
    doi = "10.1007/JHEP08(2013)090",
    journal = "JHEP",
    volume = "08",
    pages = "090",
    year = "2013"
}

@article{Ahmadain:2022eso,
    author = "Ahmadain, Amr and Wall, Aron C.",
    title = "{Off-shell strings II: Black hole entropy}",
    eprint = "2211.16448",
    archivePrefix = "arXiv",
    primaryClass = "hep-th",
    doi = "10.21468/SciPostPhys.17.1.006",
    journal = "SciPost Phys.",
    volume = "17",
    number = "1",
    pages = "006",
    year = "2024"
}

@article{Ahmadain:2025pox,
    author = "Ahmadain, Amr and Yang, Ming",
    title = "{Strings at the Tip of the Cone and Black Hole Entropy From the Worldsheet: Part I}",
    eprint = "2512.00637",
    archivePrefix = "arXiv",
    primaryClass = "hep-th",
    month = "11",
    year = "2025"
}

@article{Kazakov:1995ae,
    author = "Kazakov, Vladimir A. and Staudacher, Matthias and Wynter, Thomas",
    title = "{Character expansion methods for matrix models of dually weighted graphs}",
    eprint = "hep-th/9502132",
    archivePrefix = "arXiv",
    reportNumber = "LPTENS-95-9",
    doi = "10.1007/BF02101902",
    journal = "Commun. Math. Phys.",
    volume = "177",
    pages = "451--468",
    year = "1996"
}

@article{Maldacena:2018vsr,
    author = "Maldacena, Juan and Milekhin, Alexey",
    title = "{To gauge or not to gauge?}",
    eprint = "1802.00428",
    archivePrefix = "arXiv",
    primaryClass = "hep-th",
    doi = "10.1007/JHEP04(2018)084",
    journal = "JHEP",
    volume = "04",
    pages = "084",
    year = "2018"
}

@article{guionnet2007second,
  title={Second order asymptotics for matrix models},
  author={Guionnet, Alice and Maurel-Segala, Edouard},
  journal={The Annals of Probability},
  volume={35},
  number={6},
  pages={2160--2212},
  year={2007},
  publisher={Institute of Mathematical Statistics},
  doi={10.1214/009117907000000141},
  url={https://doi.org/10.1214/009117907000000141}
}

@article{Guo:2021blh,
    author = "Guo, Bin and Hughes, Marcel R. R. and Mathur, Samir D. and Mehta, Madhur",
    title = "{Contrasting the fuzzball and wormhole paradigms for black holes}",
    eprint = "2111.05295",
    archivePrefix = "arXiv",
    primaryClass = "hep-th",
    doi = "10.3906/fiz-2111-13",
    journal = "Turk. J. Phys.",
    volume = "45",
    number = "6",
    pages = "281--365",
    year = "2021"
}

@article{Jiang:2020cqo,
    author = "Jiang, Yikun and Kim, Manki and Wong, Gabriel",
    title = "{Entanglement entropy and edge modes in topological string theory. Part II. The dual gauge theory story}",
    eprint = "2012.13397",
    archivePrefix = "arXiv",
    primaryClass = "hep-th",
    doi = "10.1007/JHEP10(2021)202",
    journal = "JHEP",
    volume = "10",
    pages = "202",
    year = "2021"
}

@article{Donnelly:2020teo,
    author = "Donnelly, William and Jiang, Yikun and Kim, Manki and Wong, Gabriel",
    title = "{Entanglement entropy and edge modes in topological string theory. Part I. Generalized entropy for closed strings}",
    eprint = "2010.15737",
    archivePrefix = "arXiv",
    primaryClass = "hep-th",
    doi = "10.1007/JHEP10(2021)201",
    journal = "JHEP",
    volume = "10",
    pages = "201",
    year = "2021"
}

@article{Usatyuk:2024isz,
    author = "Usatyuk, Mykhaylo and Zhao, Ying",
    title = "{Closed universes, factorization, and ensemble averaging}",
    eprint = "2403.13047",
    archivePrefix = "arXiv",
    primaryClass = "hep-th",
    doi = "10.1007/JHEP02(2025)052",
    journal = "JHEP",
    volume = "02",
    pages = "052",
    year = "2025"
}

@article{Penington:2019kki,
    author = "Penington, Geoff and Shenker, Stephen H. and Stanford, Douglas and Yang, Zhenbin",
    title = "{Replica wormholes and the black hole interior}",
    eprint = "1911.11977",
    archivePrefix = "arXiv",
    primaryClass = "hep-th",
    doi = "10.1007/JHEP03(2022)205",
    journal = "JHEP",
    volume = "03",
    pages = "205",
    year = "2022"
}

@article{Saad:2021uzi,
    author = "Saad, Phil and Shenker, Stephen H. and Yao, Shunyu",
    title = "{Comments on wormholes and factorization}",
    eprint = "2107.13130",
    archivePrefix = "arXiv",
    primaryClass = "hep-th",
    doi = "10.1007/JHEP10(2024)076",
    journal = "JHEP",
    volume = "10",
    pages = "076",
    year = "2024"
}

@article{Almheiri:2019qdq,
    author = "Almheiri, Ahmed and Hartman, Thomas and Maldacena, Juan and Shaghoulian, Edgar and Tajdini, Amirhossein",
    title = "{Replica Wormholes and the Entropy of Hawking Radiation}",
    eprint = "1911.12333",
    archivePrefix = "arXiv",
    primaryClass = "hep-th",
    doi = "10.1007/JHEP05(2020)013",
    journal = "JHEP",
    volume = "05",
    pages = "013",
    year = "2020"
}

@article{Harlow:2025pvj,
    author = "Harlow, Daniel and Usatyuk, Mykhaylo and Zhao, Ying",
    title = "{Quantum mechanics and observers for gravity in a closed universe}",
    eprint = "2501.02359",
    archivePrefix = "arXiv",
    primaryClass = "hep-th",
    reportNumber = "MIT-CTP/5824",
    doi = "10.1007/JHEP02(2026)108",
    journal = "JHEP",
    volume = "02",
    pages = "108",
    year = "2026"
}

@article{Iliesiu:2024cnh,
    author = "Iliesiu, Luca V. and Levine, Adam and Lin, Henry W. and Maxfield, Henry and Mezei, M{\'a}rk",
    title = "{On the non-perturbative bulk Hilbert space of JT gravity}",
    eprint = "2403.08696",
    archivePrefix = "arXiv",
    primaryClass = "hep-th",
    doi = "10.1007/JHEP10(2024)220",
    journal = "JHEP",
    volume = "10",
    pages = "220",
    year = "2024"
}

@article{DiUbaldo:2026rly,
    author = "Di Ubaldo, Gabriele and Iliesiu, Luca V. and Lin, Henry W. and Yan, Cynthia",
    title = "{Positivity of the gravitational path integral implies the axionic weak gravity conjecture}",
    eprint = "2605.05305",
    archivePrefix = "arXiv",
    primaryClass = "hep-th",
    reportNumber = "RIKEN-iTHEMS-Report-26",
    month = "5",
    year = "2026"
}

@article{Maldacena:2026jqd,
    author = "Maldacena, Juan and Maloney, Alexander and McPeak, Brian",
    title = "{Wormholes and the imaginary distance bound}",
    eprint = "2605.05336",
    archivePrefix = "arXiv",
    primaryClass = "hep-th",
    month = "5",
    year = "2026"
}

@article{Hampapura:2020hfg,
    author = "Hampapura, Harsha R. and Harper, Jonathan and Lawrence, Albion",
    title = "{Target space entanglement in Matrix Models}",
    eprint = "2012.15683",
    archivePrefix = "arXiv",
    primaryClass = "hep-th",
    reportNumber = "BRX-TH-6658",
    doi = "10.1007/JHEP10(2021)231",
    journal = "JHEP",
    volume = "10",
    pages = "231",
    year = "2021"
}

@article{Das:2020jhy,
    author = "Das, Sumit R. and Kaushal, Anurag and Mandal, Gautam and Trivedi, Sandip P.",
    title = "{Bulk Entanglement Entropy and Matrices}",
    eprint = "2004.00613",
    archivePrefix = "arXiv",
    primaryClass = "hep-th",
    reportNumber = "TIFR/TH/20-8",
    doi = "10.1088/1751-8121/abafe4",
    journal = "J. Phys. A",
    volume = "53",
    number = "44",
    pages = "444002",
    year = "2020"
}

@article{Hartnoll:2015fca,
    author = "Hartnoll, Sean A. and Mazenc, Edward",
    title = "{Entanglement entropy in two dimensional string theory}",
    eprint = "1504.07985",
    archivePrefix = "arXiv",
    primaryClass = "hep-th",
    doi = "10.1103/PhysRevLett.115.121602",
    journal = "Phys. Rev. Lett.",
    volume = "115",
    number = "12",
    pages = "121602",
    year = "2015"
}

@article{Gautam:2022akq,
    author = "Gautam, Vaibhav and Hanada, Masanori and Jevicki, Antal and Peng, Cheng",
    title = "{Matrix entanglement}",
    eprint = "2204.06472",
    archivePrefix = "arXiv",
    primaryClass = "hep-th",
    reportNumber = "DMUS-MP-22/03",
    doi = "10.1007/JHEP01(2023)003",
    journal = "JHEP",
    volume = "01",
    pages = "003",
    year = "2023"
}

@article{Das:2022nxo,
    author = "Das, Sumit R. and Jevicki, Antal and Zheng, Junjie",
    title = "{Finiteness of entanglement entropy in collective field theory}",
    eprint = "2209.04880",
    archivePrefix = "arXiv",
    primaryClass = "hep-th",
    doi = "10.1007/JHEP12(2022)052",
    journal = "JHEP",
    volume = "12",
    pages = "052",
    year = "2022"
}

@article{Frenkel:2021yql,
    author = "Frenkel, Alexander and Hartnoll, Sean A.",
    title = "{Entanglement in the Quantum Hall Matrix Model}",
    eprint = "2111.05967",
    archivePrefix = "arXiv",
    primaryClass = "hep-th",
    doi = "10.1007/JHEP05(2022)130",
    journal = "JHEP",
    volume = "05",
    pages = "130",
    year = "2022"
}

@article{Frenkel:2023aft,
    author = "Frenkel, Alexander and Hartnoll, Sean A.",
    title = "{Emergent area laws from entangled matrices}",
    eprint = "2301.01325",
    archivePrefix = "arXiv",
    primaryClass = "hep-th",
    doi = "10.1007/JHEP05(2023)084",
    journal = "JHEP",
    volume = "05",
    pages = "084",
    year = "2023"
}

@article{Han:2019wue,
    author = "Han, Xizhi and Hartnoll, Sean A.",
    title = "{Deep Quantum Geometry of Matrices}",
    eprint = "1906.08781",
    archivePrefix = "arXiv",
    primaryClass = "hep-th",
    doi = "10.1103/PhysRevX.10.011069",
    journal = "Phys. Rev. X",
    volume = "10",
    number = "1",
    pages = "011069",
    year = "2020"
}

@article{Fliss:2025omb,
    author = "Fliss, Jackson R. and Frenkel, Alexander",
    title = "{Matrix Quantum Mechanics and Entanglement Entropy: A Review}",
    eprint = "2512.03163",
    archivePrefix = "arXiv",
    primaryClass = "hep-th",
    doi = "10.3390/e28010058",
    journal = "Entropy",
    volume = "28",
    number = "1",
    pages = "58",
    year = "2026"
}

@article{Liu:2020jsv,
    author = "Liu, Hong and Vardhan, Shreya",
    title = "{Entanglement Entropies of Equilibrated Pure States in Quantum Many-Body Systems and Gravity}",
    eprint = "2008.01089",
    archivePrefix = "arXiv",
    primaryClass = "hep-th",
    reportNumber = "MIT-CTP/5227",
    doi = "10.1103/PRXQuantum.2.010344",
    journal = "PRX Quantum",
    volume = "2",
    number = "1",
    pages = "010344",
    year = "2021"
}

@article{Lin:2017uzr,
    author = "Lin, Jennifer",
    title = "{Ryu-Takayanagi Area as an Entanglement Edge Term}",
    eprint = "1704.07763",
    archivePrefix = "arXiv",
    primaryClass = "hep-th",
    month = "4",
    year = "2017"
}

@article{Mertens:2022ujr,
    author = "Mertens, Thomas G. and Sim{\'o}n, Joan and Wong, Gabriel",
    title = "{A proposal for 3d quantum gravity and its bulk factorization}",
    eprint = "2210.14196",
    archivePrefix = "arXiv",
    primaryClass = "hep-th",
    doi = "10.1007/JHEP06(2023)134",
    journal = "JHEP",
    volume = "06",
    pages = "134",
    year = "2023"
}

@article{Kazakov:2000pm,
    author = "Kazakov, Vladimir and Kostov, Ivan K. and Kutasov, David",
    title = "{A Matrix model for the two-dimensional black hole}",
    eprint = "hep-th/0101011",
    archivePrefix = "arXiv",
    reportNumber = "SACLAY-SPH-T-00-123, LPTHENS-00-32, EFI-2000-29",
    doi = "10.1016/S0550-3213(01)00606-X",
    journal = "Nucl. Phys. B",
    volume = "622",
    pages = "141--188",
    year = "2002"
}

@article{Furuuchi:2006st,
    author = "Furuuchi, Kazuyuki",
    title = "{Matrix Model For Polyakov Loops, String Field Theory In The Temporal Gauge, And Winding String Condensation In Anti-de Sitter Space}",
    eprint = "hep-th/0608108",
    archivePrefix = "arXiv",
    month = "8",
    year = "2006"
}

@article{Aharony:2006th,
    author = "Aharony, Ofer and Komargodski, Zohar and Razamat, Shlomo S.",
    title = "{On the worldsheet theories of strings dual to free large N gauge theories}",
    eprint = "hep-th/0602226",
    archivePrefix = "arXiv",
    reportNumber = "WIS-03-06-FEB-DPP",
    doi = "10.1088/1126-6708/2006/05/016",
    journal = "JHEP",
    volume = "05",
    pages = "016",
    year = "2006"
}

@article{Geng:2021hlu,
    author = "Geng, Hao and Karch, Andreas and Perez-Pardavila, Carlos and Raju, Suvrat and Randall, Lisa and Riojas, Marcos and Shashi, Sanjit",
    title = "{Inconsistency of islands in theories with long-range gravity}",
    eprint = "2107.03390",
    archivePrefix = "arXiv",
    primaryClass = "hep-th",
    doi = "10.1007/JHEP01(2022)182",
    journal = "JHEP",
    volume = "01",
    pages = "182",
    year = "2022"
}

@article{Antonini:2025sur,
    author = "Antonini, Stefano and Chen, Chang-Han and Maxfield, Henry and Penington, Geoff",
    title = "{An apologia for islands}",
    eprint = "2506.04311",
    archivePrefix = "arXiv",
    primaryClass = "hep-th",
    doi = "10.1007/JHEP10(2025)034",
    journal = "JHEP",
    volume = "10",
    pages = "034",
    year = "2025"
}

@article{Asplund:2008xd,
    author = "Asplund, Curtis T. and Berenstein, David",
    title = "{Small AdS black holes from SYM}",
    eprint = "0809.0712",
    archivePrefix = "arXiv",
    primaryClass = "hep-th",
    doi = "10.1016/j.physletb.2009.02.043",
    journal = "Phys. Lett. B",
    volume = "673",
    pages = "264--267",
    year = "2009"
}

@article{Gross:1980he,
    author = "Gross, D. J. and Witten, Edward",
    title = "{Possible Third Order Phase Transition in the Large N Lattice Gauge Theory}",
    doi = "10.1103/PhysRevD.21.446",
    journal = "Phys. Rev. D",
    volume = "21",
    pages = "446--453",
    year = "1980"
}

@article{Wadia:1980cp,
    author = "Wadia, Spenta R.",
    title = "{$N$ = Infinity Phase Transition in a Class of Exactly Soluble Model Lattice Gauge Theories}",
    reportNumber = "EFI-80/15-CHICAGO",
    doi = "10.1016/0370-2693(80)90353-6",
    journal = "Phys. Lett. B",
    volume = "93",
    pages = "403--410",
    year = "1980"
}

@article{Aharony:2005bq,
    author = "Aharony, Ofer and Marsano, Joseph and Minwalla, Shiraz and Papadodimas, Kyriakos and Van Raamsdonk, Mark",
    title = "{A First order deconfinement transition in large N Yang-Mills theory on a small S**3}",
    eprint = "hep-th/0502149",
    archivePrefix = "arXiv",
    reportNumber = "WIS-03-05-JAN-DPP",
    doi = "10.1103/PhysRevD.71.125018",
    journal = "Phys. Rev. D",
    volume = "71",
    pages = "125018",
    year = "2005"
}

@article{Alvarez-Gaume:2005dvb,
    author = "Alvarez-Gaume, Luis and Gomez, Cesar and Liu, Hong and Wadia, Spenta",
    title = "{Finite temperature effective action, AdS(5) black holes, and 1/N expansion}",
    eprint = "hep-th/0502227",
    archivePrefix = "arXiv",
    reportNumber = "CERN-PH-TH-2004-251, IFT-05-11, MIT-CTP-3591, TIFR-TH-05-03, CERN-PH-TH-04-251",
    doi = "10.1103/PhysRevD.71.124023",
    journal = "Phys. Rev. D",
    volume = "71",
    pages = "124023",
    year = "2005"
}

@article{CugliandoloGaussian2004,
    author = "Cugliandolo, Leticia F. and Lozano, Gustavo S. and Moreno, Enrique F. and Schaposnik, Fidel A.",
    title = "{A note on Gaussian integrals over paragrassmann variables}",
    eprint = "hep-th/0209172",
    archivePrefix = "arXiv",
    doi = "10.1142/S0217751X04018506",
    journal = "Int. J. Mod. Phys. A",
    volume = "19",
    pages = "1705--1714",
    year = "2004"
}

@article{Kawahara:2007ib,
    author = "Kawahara, Naoyuki and Nishimura, Jun and Takeuchi, Shingo",
    title = "{High temperature expansion in supersymmetric matrix quantum mechanics}",
    eprint = "0710.2188",
    archivePrefix = "arXiv",
    primaryClass = "hep-th",
    reportNumber = "KEK-TH-1193",
    doi = "10.1088/1126-6708/2007/12/103",
    journal = "JHEP",
    volume = "12",
    pages = "103",
    year = "2007"
}

@article{Kawahara:2007fn,
    author = "Kawahara, Naoyuki and Nishimura, Jun and Takeuchi, Shingo",
    title = "{Phase structure of matrix quantum mechanics at finite temperature}",
    eprint = "0706.3517",
    archivePrefix = "arXiv",
    primaryClass = "hep-th",
    reportNumber = "KEK-TH-1160",
    doi = "10.1088/1126-6708/2007/10/097",
    journal = "JHEP",
    volume = "10",
    pages = "097",
    year = "2007"
}

@article{Donnelly:2016auv,
    author = "Donnelly, William and Freidel, Laurent",
    title = "{Local subsystems in gauge theory and gravity}",
    eprint = "1601.04744",
    archivePrefix = "arXiv",
    primaryClass = "hep-th",
    doi = "10.1007/JHEP09(2016)102",
    journal = "JHEP",
    volume = "09",
    pages = "102",
    year = "2016"
}

@article{Banks:1997hz,
    author = "Banks, Tom and Fischler, W. and Klebanov, Igor R. and Susskind, Leonard",
    title = "{Schwarzschild black holes from matrix theory}",
    eprint = "hep-th/9709091",
    archivePrefix = "arXiv",
    reportNumber = "PUPT-1719, UTTG-24-97",
    doi = "10.1103/PhysRevLett.80.226",
    journal = "Phys. Rev. Lett.",
    volume = "80",
    pages = "226--229",
    year = "1998"
}

@article{Berkowitz:2016znt,
    author = "Berkowitz, Evan and Hanada, Masanori and Maltz, Jonathan",
    title = "{Chaos in Matrix Models and Black Hole Evaporation}",
    eprint = "1602.01473",
    archivePrefix = "arXiv",
    primaryClass = "hep-th",
    reportNumber = "UCB-PTH-16-01, SU-ITP-16-02, LLNL-JRNL-681857, UCB-PTH-16/01, SU-ITP-16/02, YITP-16-5",
    doi = "10.1103/PhysRevD.94.126009",
    journal = "Phys. Rev. D",
    volume = "94",
    number = "12",
    pages = "126009",
    year = "2016"
}

@article{Eberhardt:2021jvj,
    author = "Eberhardt, Lorenz",
    title = "{Summing over Geometries in String Theory}",
    eprint = "2102.12355",
    archivePrefix = "arXiv",
    primaryClass = "hep-th",
    doi = "10.1007/JHEP05(2021)233",
    journal = "JHEP",
    volume = "05",
    pages = "233",
    year = "2021"
}

@article{Thavanesan:2025ibm,
    author = "Thavanesan, Ayngaran and Wall, Aron C.",
    title = "{Kosmic Field Theories: towards holographic duals for unitary string cosmologies}",
    eprint = "2510.21701",
    archivePrefix = "arXiv",
    primaryClass = "hep-th",
    doi = "10.1007/JHEP06(2026)234",
    journal = "JHEP",
    volume = "06",
    pages = "234",
    year = "2026"
}

@article{Hayden:2007cs,
    author = "Hayden, Patrick and Preskill, John",
    title = "{Black holes as mirrors: Quantum information in random subsystems}",
    eprint = "0708.4025",
    archivePrefix = "arXiv",
    primaryClass = "hep-th",
    reportNumber = "CALT-68-2659, CALT-68-2659",
    doi = "10.1088/1126-6708/2007/09/120",
    journal = "JHEP",
    volume = "09",
    pages = "120",
    year = "2007"
}

@article{Geng:2025efs,
    author = "Geng, Hao and Hung, Ling-Yan and Jiang, Yikun",
    title = "{It from ETH: Multi-interval Entanglement and Replica Wormholes from Large-$c$ BCFT Ensemble}",
    eprint = "2505.20385",
    archivePrefix = "arXiv",
    primaryClass = "hep-th",
    month = "5",
    year = "2025"
}

@article{DiFrancesco:1993cyw,
    author = "Di Francesco, P. and Ginsparg, Paul H. and Zinn-Justin, Jean",
    title = "{2-D Gravity and random matrices}",
    eprint = "hep-th/9306153",
    archivePrefix = "arXiv",
    reportNumber = "LA-UR-93-1722, SACLAY-SPH-T-93-061",
    doi = "10.1016/0370-1573(94)00084-G",
    journal = "Phys. Rept.",
    volume = "254",
    pages = "1--133",
    year = "1995"
}

@article{Jafferis:2021ywg,
    author = "Jafferis, Daniel Louis and Schneider, Elliot",
    title = "{Stringy ER = EPR}",
    eprint = "2104.07233",
    archivePrefix = "arXiv",
    primaryClass = "hep-th",
    doi = "10.1007/JHEP10(2022)195",
    journal = "JHEP",
    volume = "10",
    pages = "195",
    year = "2022"
}

@article{Gesteau:2024gzf,
    author = "Gesteau, Elliott and Marcolli, Matilde and McNamara, Jacob",
    title = "{Wormhole Renormalization: The gravitational path integral, holography, and a gauge group for topology change}",
    eprint = "2407.20324",
    archivePrefix = "arXiv",
    primaryClass = "hep-th",
    month = "7",
    year = "2024"
}

@article{Martinec:2026wuu,
    author = "Martinec, Emil J.",
    title = "{The black hole S-matrix in gauge/gravity duality}",
    eprint = "2607.18393",
    archivePrefix = "arXiv",
    primaryClass = "hep-th",
    month = "7",
    year = "2026"
}

@article{KomatsuBMN2024,
    author = "Komatsu, Shota and Martina, Adrien and Penedones, Jo{\~a}o and Suchel, No{\'e} and Vuignier, Antoine and Zhao, Xiang",
    title = "{Gravity from quantum mechanics of finite matrices}",
    eprint = "2401.16471",
    archivePrefix = "arXiv",
    primaryClass = "hep-th",
    year = "2024"
}

@article{Saad:2021rcu,
    author = "Saad, Phil and Shenker, Stephen H. and Stanford, Douglas and Yao, Shunyu",
    title = "{Wormholes without averaging}",
    eprint = "2103.16754",
    archivePrefix = "arXiv",
    primaryClass = "hep-th",
    doi = "10.1007/JHEP09(2024)133",
    journal = "JHEP",
    volume = "09",
    pages = "133",
    year = "2024"
}
 
\end{document}